\documentclass[a4paper,12pt]{article}
\usepackage{jheppub,esint,psfrag}
\usepackage[utf8]{inputenc}

\usepackage{epsfig,amssymb,amsmath,psfrag,subfigure,rotate,color,wasysym,xcolor}
\usepackage[bbgreekl]{mathbbol}

\allowdisplaybreaks 
\newcommand{\insertfig}[2]{\includegraphics[width=#1cm]{#2}}

\DeclareSymbolFontAlphabet{\mathbbm}{bbold}
\DeclareSymbolFontAlphabet{\mathbb}{AMSb}%

\def\XXint#1#2#3{{\setbox0=\hbox{$#1{#2#3}{\int}$ }
\vcenter{\hbox{$#2#3$ }}\kern-.6\wd0}}

\def \be  {\begin{equation}}
\def \ee  {\end{equation}}
\def \ba  {\begin{eqnarray}}
\def \ea  {\end{eqnarray}}
\def \baa {\begin{eqnarray*}}
\def \eaa {\end{eqnarray*}}

\def \lab #1 {\label{#1}}

\newcommand\re[1]{(\ref{#1})}
\def\d{\hbox{{d}\kern-.20em\hbox{l}}}

\def \matrix #1 {\left(\begin{array}{cc} #1 \end{array}\right)}

\def \tr {\mathop{\rm tr}\nolimits}

\newcommand \VEV [1] {\left\langle{#1}\right\rangle}
\newcommand \VVEV [1] {\left\langle\!\left\langle{#1}\right\rangle\!\right\rangle}
\newcommand \ket [1] {|{#1}\rangle}
\newcommand \bra [1] {\langle {#1}|}

\def\1{\hbox{{1}\kern-.25em\hbox{l}}}

\newcommand{\ft}[2]{{\textstyle\frac{#1}{#2}}}

\makeatletter

\input pdf-trans
\newbox\qbox
\def\usecolor#1{\csname\string\color@#1\endcsname\space}
\newcommand\bordercolor[1]{\colsplit{1}{#1}}
\newcommand\fillcolor[1]{\colsplit{0}{#1}}
\newcommand\outline[1]{\leavevmode%
  \def\maltext{#1}%
  \setbox\qbox=\hbox{\maltext}%
  \boxgs{Q q 2 Tr \thickness\space w \fillcol\space \bordercol\space}{}%
  \copy\qbox%
}
\makeatother
\newcommand\colsplit[2]{\colorlet{tmpcolor}{#2}\edef\tmp{\usecolor{tmpcolor}}%
  \def\tmpB{}\expandafter\colsplithelp\tmp\relax%
  \ifnum0=#1\relax\edef\fillcol{\tmpB}\else\edef\bordercol{\tmpC}\fi}
\def\colsplithelp#1#2 #3\relax{%
  \edef\tmpB{\tmpB#1#2 }%
  \ifnum `#1>`9\relax\def\tmpC{#3}\else\colsplithelp#3\relax\fi
}
\bordercolor{black}
\fillcolor{white}
\def\thickness{.3}

\def\1{\mathbbm{1}}

\title{Five legs @ three loops: \\
$\boldmath\mathcal{N}=4$ sYM amplitude near mass-shell}
	
\author[a]{Andrei V.~Belitsky,}
\author[b]{Leonid V.~Bork,}
\author[c,d]{Roman N.~Lee,}
\author[e,f]{Vladimir A.~Smirnov}

\affiliation[a]{Department of Physics, Arizona State University,  Tempe, AZ 85287-1504, USA}
\affiliation[b]{Dukhov Automatics Research Institute, 127055 Moscow, Russia}
\affiliation[c]{Budker Institute of Nuclear Physics, 630090 Novosibirsk, Russia}
\affiliation[d]{Joint Institute for Nuclear Research, 141980 Dubna, Russia}
\affiliation[e]{Skobeltsyn Institute of Nuclear Physics, Moscow State University, 119992 Moscow, Russia}
\affiliation[f]{Moscow Center for Fundamental and Applied Mathematics, 119992 Moscow, Russia}

\abstract{We present a three-loop analysis of the scattering amplitude of five nearly massless W-bosons in planar maximally 
supersymmetric Yang-Mills theory. The basis of the master integrals is established, making use of the unitarity-cut sewing 
technique in six-dimensional $\mathcal{N} = (1,1)$ super-Yang-Mills theory. Its dimensional reduction down to four allows us 
to generate masses for internal and external states. We descend on the special Coulomb branch of maximally supersymmetric 
Yang-Mills theory by setting all propagator masses to zero. Employing explicit expressions for all integrals that we calculated in 
a companion paper, we find a concise representation for this infrared-sensitive observable. We confirm its exponentiation, both 
for infrared and finite terms. The infrared double logarithm manifests the anticipated universality through the octagon anomalous 
dimension as its governing coefficient. Unlike our previous two-loop result, this consideration reveals that each of the three 
independent kinematic structures furnishing the amplitude possesses its own function of 't Hooft coupling.}
\begin{document}

\maketitle
\flushbottom
\setcounter{footnote} 0

\section{Introduction}

When elementary particles are involved in collisions at asymptotically large energy $E$, naively, their masses can be ignored. As $E$
is lowered, this approximation can no longer be used. However, is the inclusion of masses just a mere nuisance? Can these effects be
simply incorporated as corrections to massless results? Intuitively, this appears to be the case, but it is not accurate and, in fact,
misleading. The issue is related to the subtle anomaly-like interplay between the masses of external asymptotic states and those 
propagating in quantum loops.

In the past several years, the focus has been on the kinematical situation involving massive W-bosons interacting by means of massless
gluon exchanges \cite{Caron-Huot:2021usw,Bork:2022vat,Belitsky:2022itf,Belitsky:2023ssv,Belitsky:2024agy,Belitsky:2024dcf,Belitsky:2025bgb}.
As was observed there, the limit of near massless scattering differs drastically from the strictly massless point. These studies have been performed 
in the maximally supersymmetric Yang-Mills theory with spontaneously broken gauge symmetry \cite{Alday:2009zm}, i.e., the Coulomb branch 
of the model. In addition to being the simplest quantum field theory in four dimensions with robust control on various conditions/approximations 
employed, it enjoys the AdS/CFT duality \cite{Maldacena:1997re}. It can therefore be described at strong coupling using weak-coupling or 
classical-string methods \cite{Alday:2007hr,Alday:2007he}. We use this model as a tool to sharpen our sword, which could be later used to 
attack and defeat the strong coupling QCD problem. So far, we have used perturbation theory to uncover near massless gauge dynamics to the 
highest order achievable by modern multiloop techniques. In the present work, we push our earlier consideration of the
five-W-boson amplitude \cite{Belitsky:2025bgb} to three loops.

The task at hand requires two ingredients. First, one needs to find a basis of the so-called master integrals that determines the form of
the amplitude at three loops. Second, one needs to devise proper techniques to calculate these analytically. In this installment of our results,
we address the first problem. In the companion work \cite{BBLOS2026}, we present technical details of the methods used and solutions to all
integrals involved.

Although finding the amplitude in terms of loop integrals seems like a straightforward task, just use Feynman rules and add the resulting
diagrams together, this approach yields prohibitively long and complicated expressions that cannot be efficiently handled on computers.
This represents a bottleneck in high-loop calculations that one needs to overcome. Fortunately, a better technique is available. It is
known as the unitarity-cut sewing procedure \cite{Bern:1994zx,Bern:1994cg,Bern:2004cz,Bern:1997sc,Britto:2004nc,Cachazo:2008vp}.
As states propagating in quantum loops approach production thresholds, momentum integrals develop cuts in the complex plane of external
kinematical variables $s$. Thus, calculating their discontinuities by making use of Cutkosky rules breaks up integrands into products of on-shell
tree amplitudes. If the total amplitude does not possess additional contributions that are completely analytic in $s$, it is fixed uniquely. While
real-world theories like QCD suffer from these, the model in question is completely cut-constructible \cite{Bern:1994zx}.

However, even with the above issue settled, models with spontaneously broken gauge symmetries proliferate states by virtue of the
required decomposition of fields in terms of un- and preserved symmetry indices. Keeping track of these can be painful. Moreover, the
simplicity of the massless spinor-helicity formalism used for trees is gone when the states are massive. Here is where a high-dimensional
perspective saves the day \cite{Selivanov:1999ie,Bern:2010qa,Boels:2010mj,Craig:2011ws}. Massless maximally supersymmetric
Yang-Mills (sYM) theory can be obtained from the dimensional reduction of the ten-dimensional $\mathcal{N} = 1$ sYM down to four by
keeping momenta reciprocal to out-of-four-dimensional coordinates at zero values. The special Coulomb branch that we are currently
interested in is reached by making them nonvanishing. However, the ten-dimensional spinor-helicity formalism employs constrained
spinors \cite{Caron-Huot:2010nes} and is not particularly useful for our purposes. But there is an intermediate dimension between four and
ten that does possess an unconstrained formalism: it is six \cite{Cheung:2009dc}. The corresponding model inherits $\mathcal{N} = 2$
supersymmetry, but since it is not chiral, it is in fact $\mathcal{N} = (1,1)$ \cite{Dennen:2009vk}. Though only the tree amplitudes of multiplicity
up to (and including) six are known in it, this suffices our present needs.

Our subsequent presentation is organized as follows. In the next section, we prepare the ground for our analysis by reviewing essential
material on spinor-helicity formalism for $\mathcal{N} = (1,1)$ sYM. Then, in Sect.\ \ref{SectionCuts}, we present the calculation of the
spanning set of cuts required to fix the form of the three-loop integrand for the five-leg amplitude. The basis of the six-dimensional master
integrals is given in Sect.\ \ref{SectionBasis}. Further, we descend on the special Coulomb branch of the four-dimensional theory in Sect.\
\ref{SectionDR}. Then, relying on explicit analytic results for contributing integrals from Ref.\ \cite{BBLOS2026}, we present our findings.
Finally, we conclude.

\section{\boldmath Basics of $\mathcal{N} = (1,1)$ sYM}

We start our consideration with a compendium of a minimal set of known results about the six-dimensional $\mathcal{N} = (1,1)$ sYM
that suffice our purposes. All of the model's on-shell states, i.e., different species of scalars $\phi$, fermions $\chi, \psi$, and gluons $g$,
can be incorporated into a single CPT-self-conjugate superfield \cite{Dennen:2009vk}
\begin{align}
\label{6Dsuperfield}
{\mit\Phi} = \phi + \chi^a \eta_a + \bar{\chi}_{\dot{a}} \bar\eta^{\dot{a}} + \eta^2 \phi^\prime + \bar\eta^2 \phi^{\prime\prime}
+
g^{a}{}_{\dot{a}} \eta_a \bar\eta^{\dot{a}}
+
\psi^a \eta_a \bar\eta^2
+
\bar\psi_{\dot{a}} \bar\eta^{\dot{a}} \eta^2
+
\eta^2 \bar\eta^2 \phi^{\prime\prime\prime}
\, .
\end{align}
The field is obviously non-chiral as it is cast as a terminating series in the chiral $\eta_a$ and antichiral $\bar\eta^{\dot{a}}$
Grassmann variables. Their indices transform with respect to the non-Abelian little group SU$(2)\times$SU$(2)$. Because of
this, there is no classification with respect to helicity. This is a blessing and a curse. The former implies that there is just
one amplitude at each multiplicity, yielding a more concise classification. However, this complicates the amplitudes' form as
one loses the ability to decompose them in terms of the total helicity of particles involved in scattering, i.e., maximal helicity,
next-to-maximal helicity, etc., violation. As a result, currently, explicit tree amplitudes are available only up to six legs. As
compared to their four-dimensional counterparts, $n$-leg reduced amplitudes $\widehat{\mathcal{A}}_n$ are homogeneous
polynomials of order $(n-4)$ in $(\eta\bar\eta)$, after the extraction of the anti-/chiral charge conserving delta
function\footnote{The bosonic momentum conservation is tacitly implied everywhere through this work, but will not be explicitly displayed.}
\begin{align}
\label{SuperAmplitude}
\mathcal{A}_n
=
\widehat{\mathcal{A}}_n \,
\delta^{(4)} \left( \sum\nolimits_{i = 1}^5 Q_i \right) \delta^{(4)} \left( \sum\nolimits_{i = 1}^5 \bar{Q}_i \right)
\, .
\end{align}
The identification between the states populating the superfield \re{6Dsuperfield} and specific components of the superamplitude
\re{SuperAmplitude} is easily established by expanding its right-hand side in $\eta$ and $\bar\eta$ making use of the following
expressions for the supercharges
\begin{align}
Q_i
=
\bra{i^{a}} \eta_{i, a}
\, , \qquad
\bar{Q}_i
=
[i_{\dot{a}}| \bar\eta_i^{\dot{a}}
\, ,
\end{align}
in terms of the unconstrained six-dimensional spinors $\bra{i^{a}}$ and $[i_{\dot{a}}|$ \cite{Cheung:2009dc}. At the same time, the
six-dimensional null momenta of particles involved in the collision are given in their terms as follows
\begin{align}
P_i = \ket{i^{a}} \bra{i_a}
\, , \qquad
\bar{P}_i = |i_{\dot{a}}] [i^{\dot{a}}|
\, .
\end{align}
Here, $Q$'s and $P$'s are, respectively, rank-one and rank-two tensors in the spinor representation of the six-dimensional Lorentz group
SO$(1,5)$, inheriting the index structure from corresponding spinors. For brevity, we do not show them explicitly. Sometimes, when displaying
Dirac strings, we will equivalently write $Q$/$\bar{Q}$ as bras $\bra{Q}$/$[\bar{Q}|$ or kets $\ket{Q}$/$|\bar{Q}]$.

Only four- and five-leg color-ordered reduced amplitudes will be necessary for our construction. They are dual-conformally
covariant and read \cite{Cheung:2009dc,Dennen:2009vk,Plefka:2014fta}
\begin{align}
\label{CompactA5}
\widehat{\mathcal{A}}_4^{(0)}
= \frac{1}{S_{12} S_{23}}
\, , \qquad
\widehat{\mathcal{A}}_5^{(0)}
=
\frac{- \Omega}{S_{12} S_{23} S_{34} S_{45} S_{51}}
\, ,
\end{align}
with $S_{ij} = \ft12 \tr_4 [P_{ij}^2]$ being the six-dimensional Mandelstam invariants. There are, respectively, two and five independent
sequential-index $S_{i,i+1}$-variables for four and five particles. In the above two equations, as we already alluded to above,
$\widehat{\mathcal{A}}_4^{(0)}$ is purely bosonic, while $\widehat{\mathcal{A}}_5^{(0)}$ is bilinear in $\eta$ and $\bar\eta$. Here
$\Omega$ is determined by the SU$(4)$ inner product \cite{Huang:2011um,Plefka:2014fta}
\begin{align}
\label{Omega1}
\Omega = \ft12 \bra{B} \bar{B}] + \mbox{cc}
\end{align}
of spinors built from the supermomenta
\begin{align}
\label{Omega2}
\bra{B} = - S_{34} \bra{Q_5} + \bra{Q_{34}} \bar{P}_{34} P_5
\, , \quad
|\bar{B}]
=
- S_{51} |\bar{Q}_4] + \bar{P}_4 P_{51} |\bar{Q}_{51}]
\, .
\end{align}
Everywhere in this paper, we are using a short-hand notation for the sum of supermomenta, i.e., $P_{ij\dots} \equiv P_i + P_j + \dots$
and $Q_{ij\dots} \equiv Q_i + Q_j + \dots$.  Notice that Eq.\ \re{Omega1} can be equivalently written in terms of the other four cyclic
permutations of indices in \re{Omega2} on the shell of supercharge conservation conditions.

Calculations that follow will be enormously simplified by an educated choice of the supercomponent employed in the analysis.
Echoing considerations of Refs.\ \cite{Belitsky:2024rwv,Belitsky:2025vfc}, we use the  `singlet' amplitude $\mathcal{A}^{(0)}_5
= \mathcal{A}^{(0)}_5 (\phi'''_1,\phi'''_2, g_3, \phi_4, \phi_5)$ for this purpose. It can be extracted by setting the supercharges for
legs four and five to zero, integrating out the Grassmann variables for the first two legs, such that
\begin{align}
\label{SingletCondition}
{\rm sing}
=
\left\{
Q_1 = - \frac{Q_3 \bar{P}_2 P_1}{S_{12}}
\, ,
Q_2 = - \frac{Q_3 \bar{P}_1 P_2}{S_{12}}
\, ,
Q_4 = Q_5 = 0
\right\}
\, ,
\end{align}
with an overall Jacobian $S_{12}^2$, and, finally, integrating out $Q_3/\bar{Q}_3$ with the projector
\begin{align}
\Pi_3 = [\bar{Q}_3| P_4 \bar{P}_5 P_1 \bar{P}_2 \ket{Q_3}
\, ,
\end{align}
which selects the six-dimensional gluon for the third leg. To avoid cluttering formulas, we introduce two sets of brackets for the corresponding
operations
\begin{align}
\label{Projectors}
\VEV{\dots} = \int d^2 \eta_1 d^2 \bar\eta_1 d^2 \eta_2 d^2 \bar\eta_2  {\VVEV{\dots}}
\, ,
\qquad
{\VVEV{\dots}} = \int d^2 \eta_3 d^2 \bar\eta_3 (\dots) \Pi_3
\, .
\end{align}
In this manner, we find
\begin{align}
\VEV{\mathcal{A}^{(0)}_5}
&= - \frac{S_{12} {\VVEV{\Omega_{\rm sing}}}}{S_{23} S_{34} S_{45} S_{51}}
\, ,
\end{align}
where $\Omega_{\rm sing}$ is the $\Omega$-structure evaluated on the supercharge conditions \re{SingletCondition}
\begin{align}
\Omega_{\rm sing} = \frac{S_{45}}{S_{12}} \bra{Q_3} \bar{P}_2 P_1 \bar{P}_5 P_4 | \bar{Q}_3]
\, ,
\end{align}
which upon the use of the following Grassmann integrals
\begin{align}
\int d^2 \eta_3 \ket{Q_3} \bra{Q_3} = - P_3
\, , \qquad
\int d^2 \bar\eta_3 |\bar{Q}_3] [ \bar{Q}_3| = - \bar{P}_3
\, ,
\end{align}
generates a trace. It is calculated to yield
\begin{align}
{\VVEV{\Omega^{\rm sing}}}
&
= \frac{S_{45}}{S_{12}} \tr_4 [\bar{P}_3 P_4 \bar{P}_5 P_1 \bar{P}_2 P_3 \bar{P}_2 P_1 \bar{P}_5 P_4]
\nonumber\\
&
= 2 S_{23} S_{34} S_{45}^2 S_{51}
\, .
\end{align}

A few important points need to be emphasized about our extraction procedure. First, all expressions are covariant in the six-dimensional
space-time and corresponding inner products are expressed in terms of Madelstam invariants $S_{i,i+1}$. This implies that this formalism
is not sensitive to the so-called $\mu$-terms \cite{Bern:2002tk}, which correspond to inner products of momentum components in extra
dimensions on top of the conventional four. To detect them, we would have to break the covariance and focus on the extraction of solely
the MHV or $\overline{\rm MHV}$ amplitudes and work with six-dimensional spinors \cite{Bern:2010qa}. Instead, we forfeit their individual
extraction in favor of simplicity. This is, however, hardly a limitation to our goals since the loop-momentum integrals are strictly performed
in four-dimensional space, as they are finite, being regularized by the external legs' off-shellness. Next, since we project on a six-dimensional
gluon for the third leg, we are effectively extracting the sum of both gluon helicities from the four-dimensional point of view since
\begin{align}
\tr_4 \stackrel{\rm dim.red.}{\to} \tr_{\rm chiral} + \tr_{\rm antichiral}
\, ,
\end{align}
i.e., the sum of MHV or $\overline{\rm MHV}$ amplitudes. This wipes out the parity-odd contribution to the amplitude. The latter were shown
not to affect the parity-even part of amplitudes in the massless limit \cite{Spradlin:2008uu}, and we will assume that this pattern continues off
the mass shell.

\section{Iterated cuts}
\label{SectionCuts}

\begin{figure}[t]
\begin{center}
\mbox{
\begin{picture}(0,250)(230,0)
\put(0,-10){\insertfig{16}{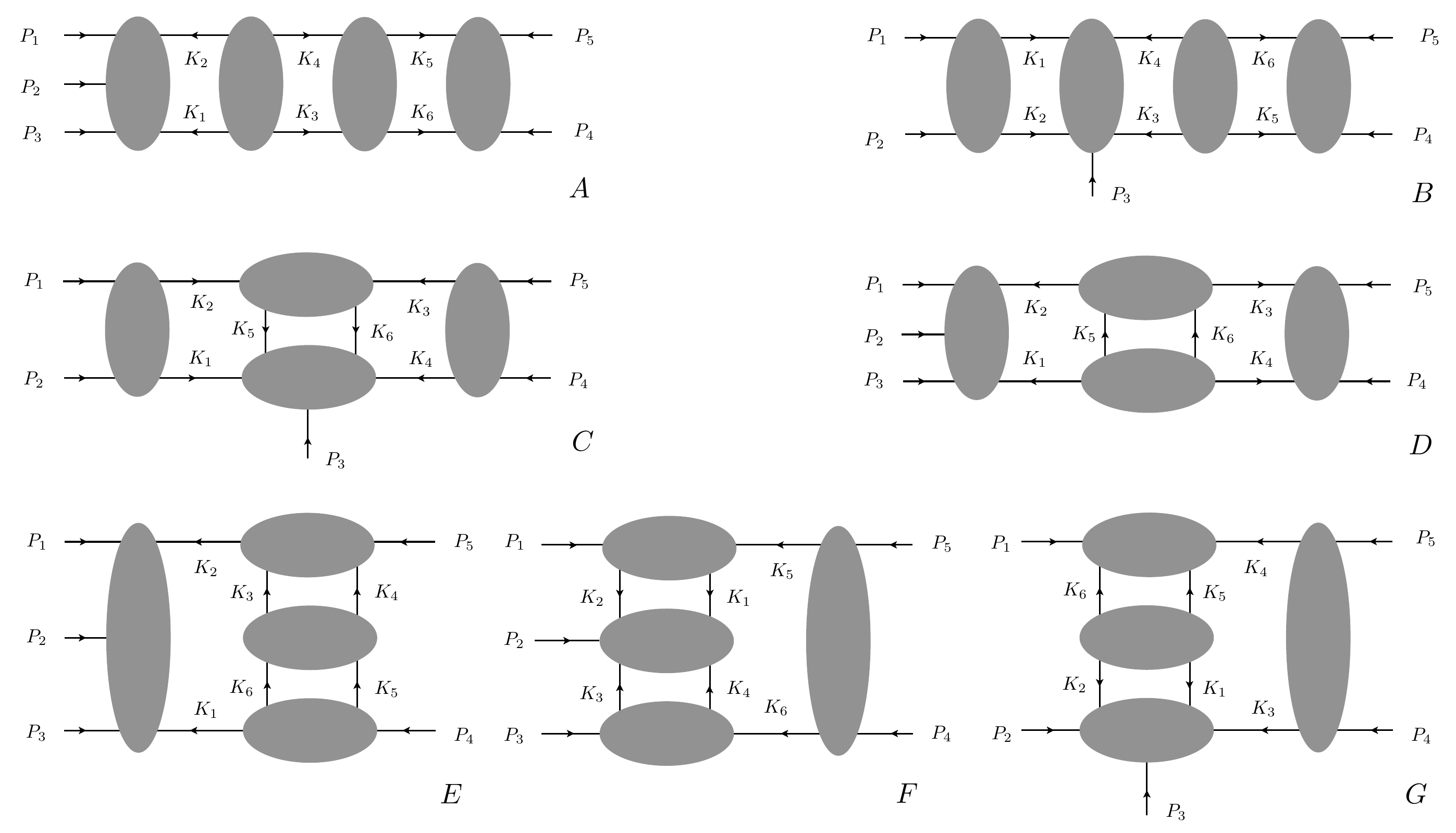}}
\end{picture}
}
\end{center}
\caption{\label{CutsGraphs} The spanning set of iterated unitarity cuts required for reconstruction of the three-loop
five-leg amplitude. Cut D does not bring anything new compared to cut A.}
\end{figure}

Now, we are in a position to find the basis of independent integrals at three loops. We will not conjecture their form, but rather allow for
the cut-sewing procedure to determine them for us. The only assumption we have to make at this stage is that a basis of the master
integrals\footnote{Our amplitude is represented as a linear combination of basic (master) integrals, similarly to how Feynman integrals of
a given family are expressed in terms of the corresponding master integrals identified with the help of integration-by-parts relations.}
${\mathcal I}_\alpha$ does exist. We decompose the three-loop amplitude as follows
\begin{align}
\label{3LBasisDecomposition}
\mathcal{A}^{(3)}_5
=
\mathcal{A}^{(0)}_5 \sum_\alpha c_\alpha {\mathcal I}_\alpha
\, ,
\end{align}
with unknown (at this stage) coefficients $c_\alpha$. The latter are multimonomials in Madelstam invariants with rational coefficients.

Next, we compute the left-hand side of \re{3LBasisDecomposition} on a spanning set of cuts shown in Fig.\ \ref{CutsGraphs}.
Of course, neither the coefficients $c_\alpha$ nor the basis elements ${\mathcal I}_\alpha$ depend on the spin degrees of freedom
of external states. However, the Grassmann integrations essentially become trivial with our choice of the `singlet' supercomponent
introduced above. We obtain the following expressions on unitarity cuts displayed in Fig.\ \ref{CutsGraphs} before the final projection
on the six-dimensional gluon\footnote{In our conventions, the overall minus sign comes from the product of $i$'s of the six cut 
propagators.}:
\begin{alignat}{2}
\VEV{ \mathcal{A}^{(3)}_5|_{\rm cut-A} }
&=
\frac{- S_{12}^2 S_{45}^3}{S_{K_2 K_4} S_{K_4, -K_5} S_{5, K_5}}
&&
{\VVEV{
\mathcal{A}^{(0)}_5 (P_1, P_2, P_3, K_1, K_2) |_{{\rm conds}_{\rm A}}
}}
\, , \nonumber\\
\VEV{ \mathcal{A}^{(3)}_5|_{\rm cut-B} }
&=
\frac{- S_{12}^3 S_{45}^2}{S_{1, -K_1} S_{K_4, K_6} S_{5, K_6}}
&&
{\VVEV{
\mathcal{A}^{(0)}_5 (K_1, K_2, P_3, K_3, K_4) |_{{\rm conds}_{\rm B}}
}}
\, , \nonumber\\
\VEV{ \mathcal{A}^{(3)}_5|_{\rm cut-C} }
&=
\frac{- S_{12}^3 S_{45} S_{K_2 K_3}}{S_{2, -K_1} S_{K_2, -K_5} S_{5, -K_3}}
&&
{\VVEV{
\mathcal{A}^{(0)}_5 (K_5, K_1, P_3, K_4, K_6) |_{{\rm conds}_{\rm C}}
}}
\, , \nonumber\\
\VEV{ \mathcal{A}^{(3)}_5|_{\rm cut-D} }
&=
\frac{- S_{12}^2 S_{45}^3}{S_{K_2, -K_5} S_{K_1 K_5} S_{5, K_3}}
&&
{\VVEV{
\mathcal{A}^{(0)}_5 (P_1, P_2, P_3, K_1, K_2) |_{{\rm conds}_{\rm D}}
}}
\, , \nonumber\\
\VEV{ \mathcal{A}^{(3)}_5|_{\rm cut-E} }
&=
\frac{- S_{12}^2 S_{45}^2 S_{5, -K_2}}{S_{K_2, K_3} S_{K_3, -K_6} S_{K_1K_5}}
&&
{\VVEV{
\mathcal{A}^{(0)}_5 (P_1, P_2, P_3, K_1, K_2) |_{{\rm conds}_{\rm E}}
}}
\, , \nonumber\\
\VEV{ \mathcal{A}^{(3)}_5|_{\rm cut-F} }
&=
\frac{- S_{12}^2 S_{45} S_{1, K_5} S_{3, K_6}}{S_{1, -K_2} S_{3, -K_5} S_{5, -K_5}}
&&
{\VVEV{
\mathcal{A}^{(0)}_5 (K_2, P_2, K_3, K_4, K_1) |_{{\rm conds}_{\rm F}}
}}
\, , \nonumber\\
\VEV{ \mathcal{A}^{(3)}_5|_{\rm cut-G} }
&=
\frac{- S_{12}^2 S_{45} S_{1, K_4}^2}{S_{1, K_6} S_{K_2 K_6} S_{5, -K_4}}
&&
{\VVEV{
\mathcal{A}^{(0)}_5 (K_2, P_2, P_3, K_3, K_1) |_{{\rm conds}_{\rm G}}
}}
\, .
\end{alignat}
Here, the overall factor of ratios of inner products stems from Jacobians of Grassmann integrations and denominators of the four-leg amplitudes.
The conditions imposed on the tree five-leg amplitudes arise from the solution of supercharge conservation.
They are
\begin{alignat}{2}
{\rm conds}_{\rm A}
&
=
{\rm conds}_{\rm D}
=
{\rm conds}_{\rm E}
&&
\\
&
=
\bigg\{
Q_{K_1} = Q_{K_2} = 0
\, ,
Q_{1} = - \frac{Q_3 \bar{P}_2 P_1}{S_{12}}
\, ,
Q_{2} = - \frac{Q_3 \bar{P}_1 P_2}{S_{12}}
\bigg\}
\, , && \nonumber\\
{\rm conds}_{\rm B}
&
=
\bigg\{
Q_{K_3} = Q_{K_4} = 0
\, ,
Q_{K_1} = - \frac{Q_3 \bar{K}_2 K_1}{S_{K_1 K_2}}
\, ,
Q_{K_2} = - \frac{Q_3 \bar{K}_1 K_2}{S_{K_1 K_2}}
&&
\bigg\}
\, ,
\\
{\rm conds}_{\rm C}
&
=
\bigg\{
Q_{K_4} = 0
\, ,
Q_{K_1} = - \frac{Q_3 \bar{K}_2 K_1}{S_{K_1 K_2}}
\, ,
Q_{K_5} = - \frac{Q_3 \bar{K}_1 K_2 \bar{K}_6 K_5}{S_{K_1 K_2} S_{K_5 K_6}}
\, ,
&&
\,
Q_{K_6} = - \frac{Q_3 \bar{K}_1 K_2 \bar{K}_5 K_6}{S_{K_1 K_2} S_{K_5 K_6}}
\bigg\}
\, , \\
{\rm conds}_{\rm F}
&
=
\bigg\{
Q_2 = - \frac{Q_3 \bar{P}_1 P_2}{S_{12}}
\, ,
Q_{K_3} = \frac{Q_3 \bar{K}_4 K_3}{S_{K_3 K_4}}
\, ,
Q_{K_4} = \frac{Q_3 \bar{K}_3 K_4}{S_{K_3 K_4}}
\, ,
&&
\,
Q_{K_1} = - \frac{Q_3 \bar{P}_2 P_1 \bar{K}_2 K_1}{S_{12} S_{K_1 K_2}}
\, ,
\nonumber\\
& &&
Q_{K_2} = - \frac{Q_3 \bar{P}_2 P_1 \bar{K}_1 K_2}{S_{12} S_{K_1 K_2}}
\bigg\}
\, , \\
{\rm conds}_{\rm G}
&
=
\bigg\{
Q_2 = - \frac{Q_3 \bar{P}_1 P_2}{S_{12}}
\, ,
Q_{K_3} = 0
\, ,
Q_{K_1} = - \frac{Q_3 \bar{P}_2 P_1 \bar{K}_2 K_1}{S_{12} S_{K_1 K_2}}
\, ,
&&
Q_{K_2} = - \frac{Q_3 \bar{P}_2 P_1 \bar{K}_1 K_2}{S_{12} S_{K_1 K_2}}
\bigg\}
\, .
\end{alignat}
The tree-level five-leg amplitudes evaluated on these solutions then read
\begin{align}
\mathcal{A}^{(0)}_5 (P_1, P_2, P_3, K_1, K_2) |_{{\rm conds}_{\rm A}}
&
=
\mathcal{A}^{(0)}_5 (P_1, P_2, P_3, K_1, K_2) |_{{\rm conds}_{\rm D}}
\nonumber\\
&
=
\mathcal{A}^{(0)}_5 (P_1, P_2, P_3, K_1, K_2) |_{{\rm conds}_{\rm E}}
\nonumber\\
&
=
\frac{- \Omega_{\rm A}}{S_{12} S_{23} S_{3, K_1} S_{K_1 K_2} S_{1, K_2}}
\, , \\
\mathcal{A}^{(0)}_5 (K_1, K_2, P_3, K_3, K_4) |_{{\rm conds}_{\rm B}}
&
=
\frac{- \Omega_{\rm B}}{S_{K_1 K_2} S_{3, K_2} S_{3, K_3} S_{K_3 K_4} S_{K_1 K_4}}
\, , \\
\mathcal{A}^{(0)}_5 (K_5, K_1, P_3, K_4, K_6) |_{{\rm conds}_{\rm C}}
&
=
\frac{- \Omega_{\rm C}}{S_{K_1 K_5} S_{3, K_1} S_{3, K_4} S_{K_4 K_6} S_{K_5 K_6}}
\, , \\
\mathcal{A}^{(0)}_5 (K_2, P_2, K_3, K_4, K_1) |_{{\rm conds}_{\rm F}}
&
=
\frac{- \Omega_{\rm F}}{S_{2, K_2} S_{2, K_3} S_{K_3 K_4} S_{K_1 K_4} S_{K_1 K_2}}
\, , \\
\mathcal{A}^{(0)}_5 (K_2, P_2, P_3, K_3, K_1) |_{{\rm conds}_{\rm G}}
&
=
\frac{- \Omega_{\rm G}}{S_{2, K_2} S_{23} S_{3, K_3} S_{K_1 K_3} S_{K_1 K_2}}
\, ,
\end{align}
where the $\Omega$-structures \re{Omega1} become
\begin{align}
2 \Omega_{\rm A}
&=
\frac{S_{K_1 K_2}}{S_{12}} \bra{Q_3} \bar{P}_2 P_1 \bar{K}_2 K_1 | Q_3] + {\rm cc}
\, , \\
2 \Omega_{\rm B}
&=
\frac{S_{K_3 K_4}}{S_{K_1 K_2}} \bra{Q_3} \bar{K}_2 K_1 \bar{K}_4 K_3 | Q_3] + {\rm cc}
\, , \\
2 \Omega_{\rm C}
&=
\frac{S_{K_4 K_6}}{S_{K_1 K_2}}
\bra{Q_3} \bar{K}_1 K_2
\left(
1 + \frac{\bar{K}_5 K_4}{S_{K_4 K_6}}
\right)
\bar{K_6}
\left(
K_4 + \frac{S_{3, K_4}}{S_{K_1 K_2} S_{K_5 K_6}} K_5 \bar{K}_2 K_1
\right)
| \bar{Q}_3 ]
+ {\rm cc}
\, , \\
2 \Omega_{\rm F}
&=
\frac{S_{K_1 K_2}}{S_{K_3 K_4}}
\bra{Q_3}
\left(
\frac{S_{K_3 K_4}}{S_{12} S_{K_1 K_2}} \bar{P}_2 P_1 \bar{K}_2 + \bar{K}_{34}
\right)
K_1 \bar{K}_4
\\
&\qquad\qquad\qquad\qquad\qquad\qquad\qquad\quad
\times
\left(
-
\frac{S_{K_3 K_4}}{S_{12} S_{K_1 K_2}} K_{12} \bar{P}_1 P_2
-
K_3
\right)
| \bar{Q}_3 ]
+ {\rm cc}
\, , \nonumber\\
2 \Omega_{\rm G}
&=
\frac{1}{S_{12}}
\bra{Q_3} \bar{P}_2 P_1
\left(
\frac{S_{23}}{S_{K_1 K_2}} \bar{K}_1 + \bar{P}_{23}
\right)
K_2
\left(
\bar{K}_1 K_3 + \frac{S_{3, K_3}}{S_{12}} \bar{P}_1 P_2
\right)
|\bar{Q}_3]
+
{\rm cc}
\, .
\end{align}
We remind the reader that we used the following abbreviations $P_{ij\dots} = P_i + P_j + \dots$ and
$K_{ij\dots} = K_i + K_j + \dots$ in these expressions.

Finally, the projection \re{Projectors} on the gluon yields a concise expansion in terms of six-dimensional invariants
involving external and cut loop momenta
\begin{align}
\frac{S_{12}}{S_{45}}
{\VVEV{\Omega_A}}
&
=
\ft12
\tr_4
[\bar{P}_3 P_4 \bar{P}_5 P_1 \bar{P_2} P_3 \bar{P}_2 P_1 \bar{K}_2 K_1]
-
\ft12
\tr_4
[\bar{P}_3 P_4 \bar{P}_5 P_1 \bar{P_2} P_3 \bar{K}_1 K_2 \bar{P}_1 P_2]
\nonumber\\
&
=
S_{12} S_{23} S_{34} S_{45} S_{1, K_2}
+
S_{12} S_{23} S_{45} S_{51} S_{3, K_1}
+
S_{12}^2 S_{23}^2 S_{4, -K_1}
\, , \\[2mm]
\frac{S_{12}}{S_{45}}
{\VVEV{\Omega_B}}
&
=
\ft12
\tr_4
[\bar{P}_3 P_4 \bar{P}_5 P_1 \bar{P_2} P_3 \bar{K}_2 K_1 \bar{K}_4 K_3]
-
\ft12
\tr_4
[\bar{P}_3 P_4 \bar{P}_5 P_1 \bar{P_2} P_3 \bar{K}_3 K_4  \bar{K}_1 K_2]
\nonumber\\
&
=
S_{12} S_{34} S_{45} S_{1, K_4} S_{3, K_2}
\nonumber\\
&
+
S_{12}^2 S_{23} S_{3, K_2} S_{5, -K_4}
+
S_{12} S_{23} S_{45} S_{3, K_3} S_{5, K_1}
+
S_{34} S_{45}^2 S_{1, -K_1} S_{3, K_3}
\, , \\[2mm]
S_{12}^2 S_{K_2 K_3}
{\VVEV{\Omega_C}}
&
=
\ft12
\tr_4
[\bar{P}_3 P_4 \bar{P}_5 P_1 \bar{P_2} P_3
\bar{K}_1 K_2
\left(
S_{K_4 K_6} + \bar{K}_{5} K_4
\right)
\nonumber\\
&
\qquad\qquad\qquad\qquad\qquad
\times
\bar{K}_6
\left(
S_{12} S_{K_2 K_3} K_4 + S_{3, K_4} K_5 \bar{K}_2 K_1
\right)
]
\nonumber\\
&
-
\ft12
\tr_4
[\bar{P}_3 P_4 \bar{P}_5 P_1 \bar{P_2} P_3
\left(
S_{12} S_{K_2 K_3} \bar{K}_4 + S_{3, K_4} \bar{K}_1 K_2 \bar{K}_5
\right)
\nonumber\\
&
\qquad\qquad\qquad\qquad\qquad
\times
K_6
\left(
S_{K_4 K_6} + \bar{K}_{4} K_5
\right)
\bar{K}_2 K_1
]
\nonumber\\
&
=
S_{34} S_{45}^2 S_{12} S_{3,K_1} S_{3,K_4} S_{K_2K_3} S_{1,K_5-K_2}
+
S_{23} S_{45} S_{12}^2 S_{3,K_1} S_{3,K_4} S_{K_2K_3} S_{5, K_6-K_3}
\nonumber\\
&
+
S_{34} S_{45} S_{12}^2 S_{3,K_1} S_{1, K_3} S_{K_2K_3} S_{K_4 K_6}
+
S_{23} S_{45}^2 S_{12} S_{5, K_2} S_{3,K_4} S_{K_1 K_5} S_{K_2K_3}
\nonumber\\
&
+
S_{34} S_{45} S_{12} S_{3,K_1}^2 S_{1, K_3} S_{3,K_4} S_{K_2, -K_5}
+
S_{23} S_{45} S_{12} S_{3,K_1} S_{5, K_2} S_{3,K_4}^2 S_{K_2, -K_5}
\nonumber\\
&
+
S_{34} S_{45}^3 S_{2, -K_1} S_{3,K_4} S_{K_1 K_5} S_{K_2K_3}
+
S_{23} S_{12}^3 S_{3,K_1} S_{5, -K_3} S_{K_2K_3} S_{K_4 K_6}
\nonumber\\
&
+
S_{23} S_{12}^2 S_{3,K_1}^2 S_{3,K_4} S_{5, -K_3} S_{K_2, -K_5}
+
S_{34}  S_{45}^2 S_{2, -K_1} S_{3,K_1} S_{3,K_4}^2 S_{K_2, -K_5}
\nonumber\\
&
-
S_{23} S_{34} S_{45} S_{12} S_{3,K_1} S_{3,K_4} S_{K_2K_3} S_{K_2, -K_5}
\, , \\[2mm]
S_{12}^2 S_{1,K_5} S_{3, K_6}
{\VVEV{\Omega_F}}
&
=
\ft12
\tr_4
[\bar{P}_3 P_4 \bar{P}_5 P_1 \bar{P_2} P_3
\left(
S_{K_3 K_4} \bar{P}_2 P_1 \bar{K}_2 + S_{12} S_{K_1 K_2} \bar{K}_{34}
\right)
\nonumber\\
&
\qquad\qquad\qquad\qquad\qquad
\times
K_1 \bar{K}_4
\left(
-
S_{12} S_{K_1 K_2} K_3
-
S_{K_3 K_4} K_{12} \bar{P}_1 P_2
\right)
]
\nonumber\\
&
+
\ft12
\tr_4
[\bar{P}_3 P_4 \bar{P}_5 P_1 \bar{P_2} P_3
\left(
-
S_{12} S_{K_1 K_2} \bar{K}_3
-
S_{K_3 K_4}  \bar{K}_{12} P_1 \bar{P}_2
\right)
\nonumber\\
&
\qquad\qquad\qquad\qquad\qquad
\times
K_4 \bar{K}_1
\left(
S_{K_3 K_4} P_2 \bar{P}_1 K_2 + S_{12} S_{K_1 K_2} K_{34}
\right)
]
\nonumber\\
&=
S_{12}
[
S_{12}S_{34} S_{45} S_{23,K_2}S_{2,K_3} S_{1,K_5}^2 S_{3,K_6}
+
S_{12}S_{23} S_{45} S_{2,K_2} S_{34,-K_3} S_{1,K_5}^2 S_{3,K_6}
\nonumber\\
&\qquad
+
S_{12}S_{23} S_{45} S_{51,-K_2} S_{2,K_3} S_{1,K_5} S_{3,K_6}^2
+
S_{51} S_{23} S_{45} S_{2,K_2} S_{12,K_3} S_{1,K_5} S_{3,K_6}^2
\nonumber\\
&\qquad
+
S_{12}S_{23}^2 S_{2,K_2} S_{12,K_3} S_{5,-K_5} S_{1,K_5} S_{3,K_6}
+
S_{12}^2 S_{23} S_{2,K_2} S_{3,-K_3} S_{5,-K_5} S_{1,K_5}^2
\nonumber\\
&\qquad
+
S_{12}^2 S_{23} S_{23,K_2}S_{2,K_3} S_{5,-K_5} S_{1,K_5} S_{3,K_6}
+
S_{12}S_{23}^2 S_{1,-K_2} S_{2,K_3} S_{5,-K_5} S_{3,K_6}^2
\nonumber\\
&\qquad
+
S_{51} S_{23} S_{45} S_{1,-K_2} S_{2,K_3} S_{3,K_6}^3
-
S_{12}S_{23} S_{45} S_{2,K_2} S_{2,K_3} S_{5,-K_5} S_{1,K_5} S_{3,K_6}
\nonumber\\
&\qquad
+
S_{12}S_{34} S_{45} S_{2,K_2} S_{3,-K_3} S_{1,K_5}^3
]
\, , \\[2mm]
S_{12}^2 S_{1,K_4}
{\VVEV{\Omega_G}}
&
=
\ft12
\tr_4
[\bar{P}_3 P_4 \bar{P}_5 P_1 \bar{P_2} P_3
\bar{P}_2 P_1
\left(
S_{23} \bar{K}_1 + S_{1,K_4} \bar{P}_{23}
\right)
\nonumber\\
&
\qquad\qquad\qquad\qquad\qquad\qquad\qquad\quad
\times
K_2
\left(
S_{12} \bar{K}_1 K_3 + S_{3, K_3} \bar{P}_1 P_2
\right)
]
\nonumber\\
&
-
\ft12
\tr_4
[\bar{P}_3 P_4 \bar{P}_5 P_1 \bar{P_2} P_3
\left(
S_{12} \bar{K}_3 K_1 + S_{3, K_3} \bar{P}_2 P_1
\right)
\nonumber\\
&
\qquad\qquad\qquad\qquad\qquad\qquad\qquad\quad
\times
\bar{K}_2
\left(
S_{23} K_1 + S_{1,K_4} P_{23}
\right)
\bar{P}_1 P_2
]
\nonumber\\
&=
S_{12} S_{23}
\Big[
S_{12} S_{34} S_{45} S_{1, K_4}^2 S_{K_1 K_3}
+
S_{12}^2 S_{23} S_{5, - K_4} S_{1, K_4} S_{K_1 K_3}
\nonumber\\
&\qquad\qquad
+
S_{12} S_{23}^2 S_{1,-K_2} S_{3, K_3} S_{5,-K_4}
+
S_{51} S_{45}^2 S_{2,K_2}S_{3, K_3} S_{1,K_4}
\nonumber\\
&\qquad\qquad
+
S_{23} S_{45} S_{51} S_{1,-K_2} S_{3, K_3}^2
+
S_{12} S_{23} S_{45}  S_{3, K_3} S_{1,K_4} S_{4,-K_1-K_3}
\Big]
\, .
\end{align}
Though the notations are self-explanatory, let us point out that, e.g., $S_{4,-K_1-K_3} = (P_4 - K_1 - K_3)^2$, etc.

\section{Identifying basis}
\label{SectionBasis}

Having the above unitarity cuts at our disposal, we can identify the basis of integrals in a very straightforward manner. The easiest
way to achieve this is to employ their diagrammatic representation. Below, we provide formulas for the sum of graphs for each iterated cut.
For the reader's convenience, they are displayed in a manner where they are in one-to-one correspondence with the terms appearing in
the traces. The dotted lines represent cuts in $(K_1,K_2)$ (green), $(K_3,K_4)$ (blue), and $(K_5,K_6)$ (red) channels. The numerators
are displayed explicitly above/below each Feynman graph for the master integrals in the basis. We find
\begin{align}
&
- \VEV{
\mathcal{A}^{(3)}_5|_{\rm cut-A}
}
/
\VEV{
\mathcal{A}^{(0)}_5
}
=
\frac{\ft12 S_{45} {\VVEV{\Omega_{\rm A}}}}{S_{23} S_{1,K_2} S_{3,K_1} S_{K_2 K_4} S_{K_4, -K_5} S_{5, K_5}}
\\[5mm]
&
\parbox[c][30mm][t]{170mm}{
\insertfig{17.2}{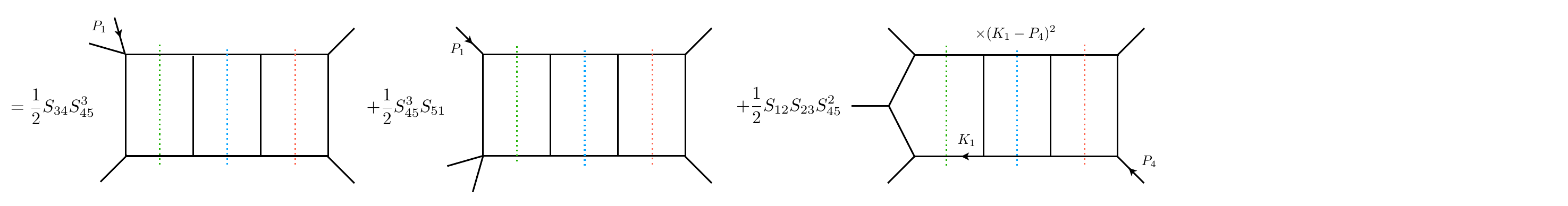}
}
\nonumber\\
&
- \VEV{
\mathcal{A}^{(3)}_5|_{\rm cut-B}
}
/
\VEV{
\mathcal{A}^{(0)}_5
}
=
\frac{\ft12 S_{12} {\VVEV{\Omega_{\rm B}}}}{S_{1,-K_1} S_{3,K_2} S_{3,K_3} S_{K_1 K_4} S_{K_4,K_6} S_{5,K_6}}
\\[5mm]
&
\parbox[c][30mm][t]{170mm}{
\insertfig{17.2}{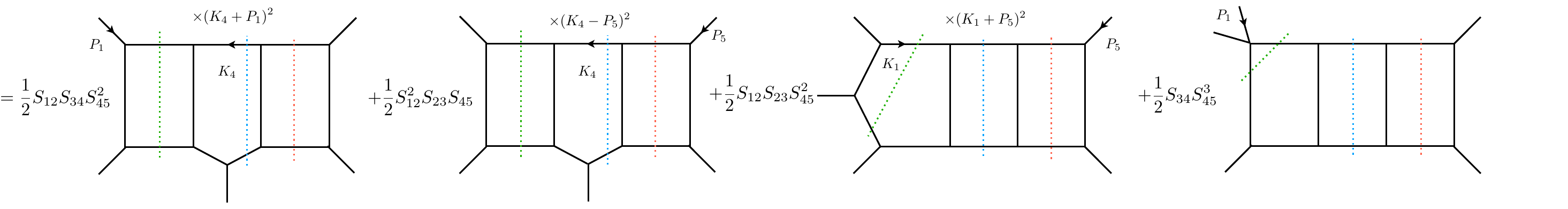}
}
\nonumber\\
&
- \VEV{
\mathcal{A}^{(3)}_5|_{\rm cut-C}
}
/
\VEV{
\mathcal{A}^{(0)}_5
}
=
\frac{\ft12 S_{12}^2 {\VVEV{\Omega_{\rm C}}}}{S_{1,-K_2} S_{K_2,-K_5} S_{5, -K_3} S_{3,K_1} S_{3,K_4} S_{K_1K_5} S_{K_4 K_6}}
\\[5mm]
&
\parbox[c][95mm][t]{170mm}{
\insertfig{17.2}{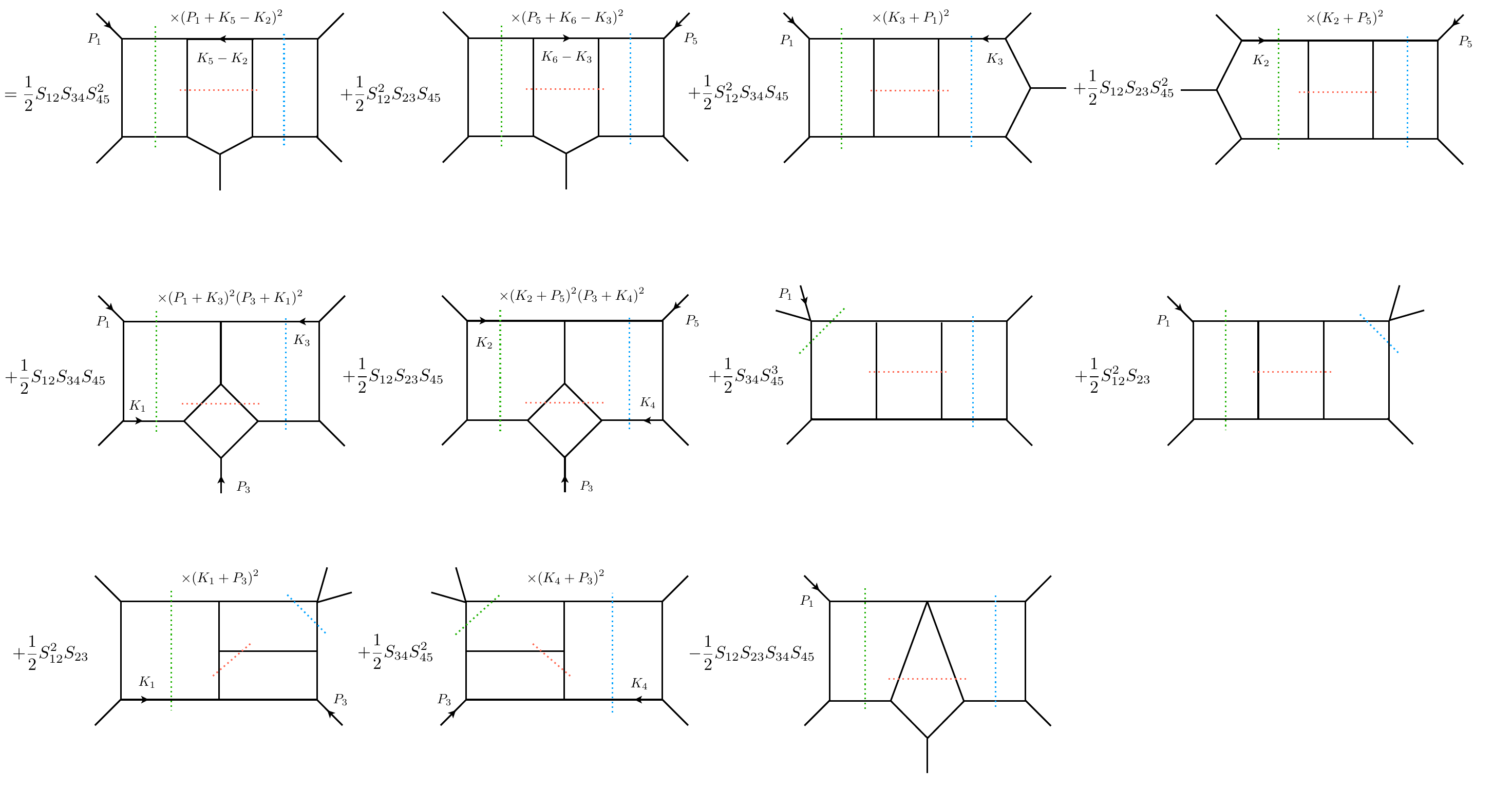}
}
\nonumber\\
&
- \VEV{
\mathcal{A}^{(3)}_5|_{\rm cut-D}
}
/
\VEV{
\mathcal{A}^{(0)}_5
}
=
\frac{\ft12 S_{45} {\VVEV{\Omega_{\rm A}}}}{S_{23} S_{1,K_2} S_{3,K_1} S_{K_2,- K_5} S_{K_1 K_5} S_{5, K_3}}
\\[5mm]
&
\parbox[c][30mm][t]{170mm}{
\insertfig{17.2}{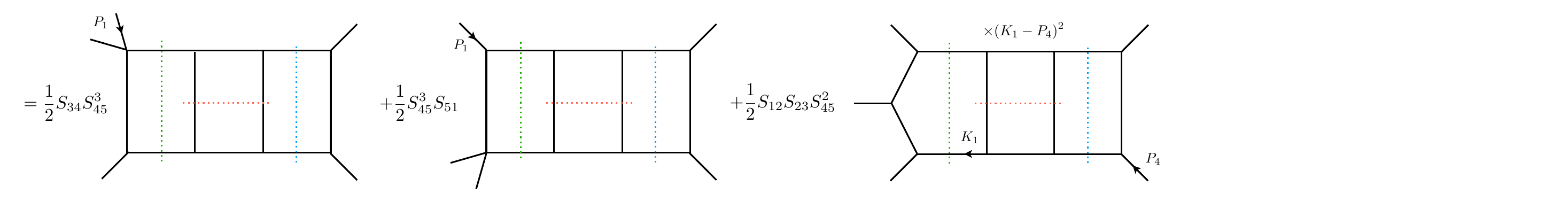}
}
\nonumber\\
&
- \VEV{
\mathcal{A}^{(3)}_5|_{\rm cut-E}
}
/
\VEV{
\mathcal{A}^{(0)}_5
}
=
\frac{\ft12 S_{12} S_{45} S_{5, -K_2}{\VVEV{\Omega_{\rm A}}}}{S_{23} S_{1,K_2} S_{3,K_1} S_{K_2,- K_3} S_{K_3, -K_6} S_{K_1 K_6}}
\\[5mm]
&
\parbox[c][30mm][t]{170mm}{
\insertfig{17.2}{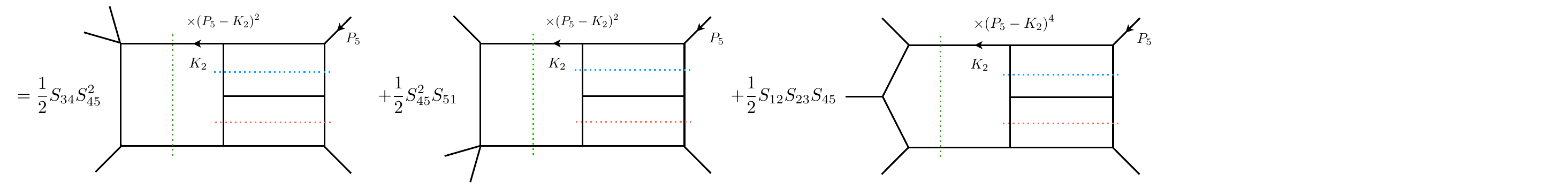}
}
\nonumber\\
&
- \VEV{
\mathcal{A}^{(3)}_5|_{\rm cut-F}
}
/
\VEV{
\mathcal{A}^{(0)}_5
}
=
\frac{\ft12 S_{12} {\VVEV{\Omega_{\rm F}}}}{S_{1,-K_2} S_{3,-K_3} S_{5,-K_5} S_{2,K_2} S_{2,K_3} S_{K_1K_4}}
\\[5mm]
&
\parbox[c][95mm][t]{170mm}{
\insertfig{17.2}{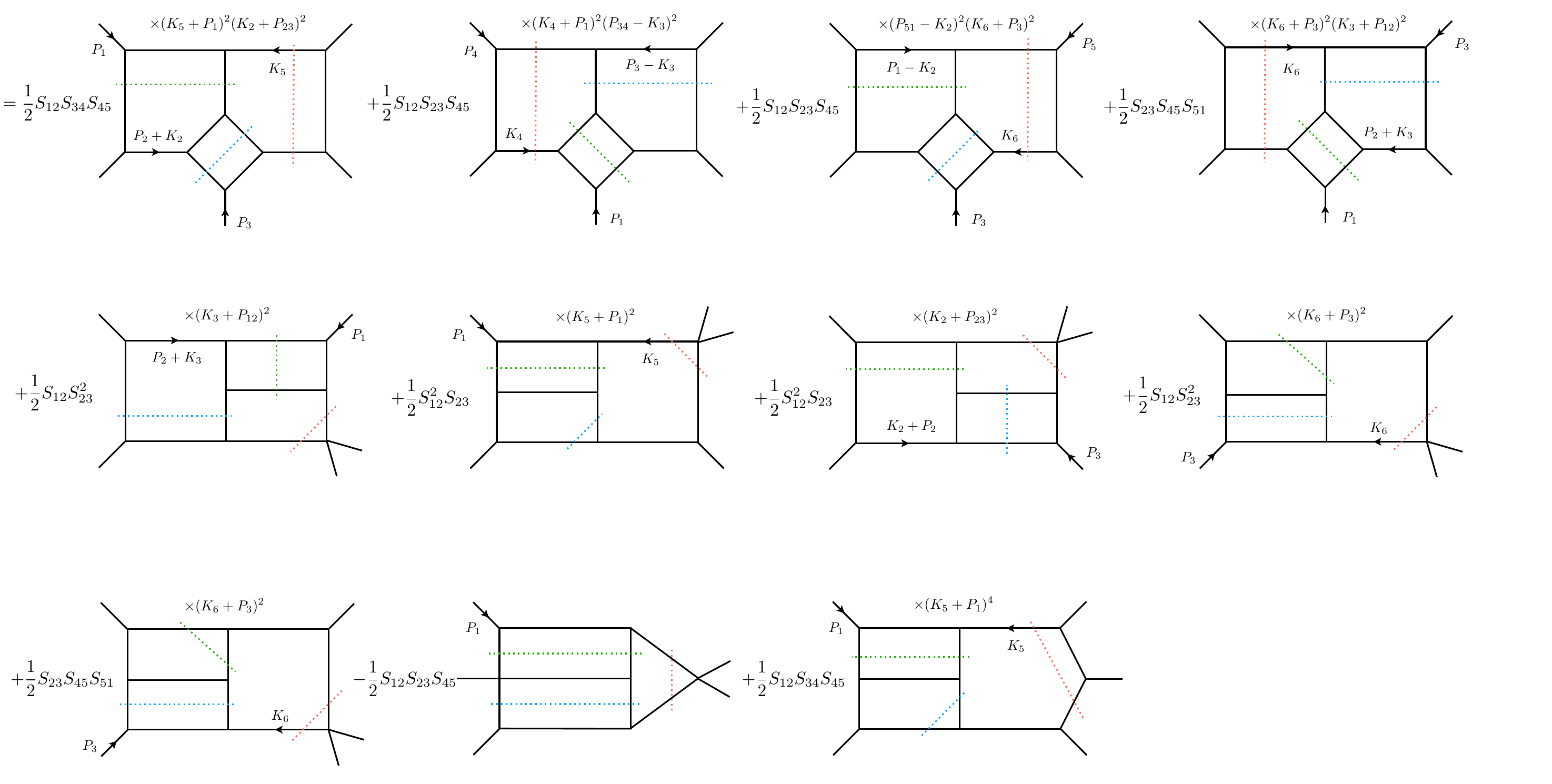}
}
\nonumber\\
&
- \VEV{
\mathcal{A}^{(3)}_5|_{\rm cut-G}
}
/
\VEV{
\mathcal{A}^{(0)}_5
}
=
\frac{\ft12 S_{12} S_{1, K_4} {\VVEV{\Omega_{\rm G}}}}{S_{23} S_{1,K_6} S_{K_2 K_6} S_{5, -K_4} S_{2, K_2} S_{3, K_3} S_{K_1 K_3}}
\\[5mm]
&
\parbox[c][60mm][t]{170mm}{
\insertfig{17.2}{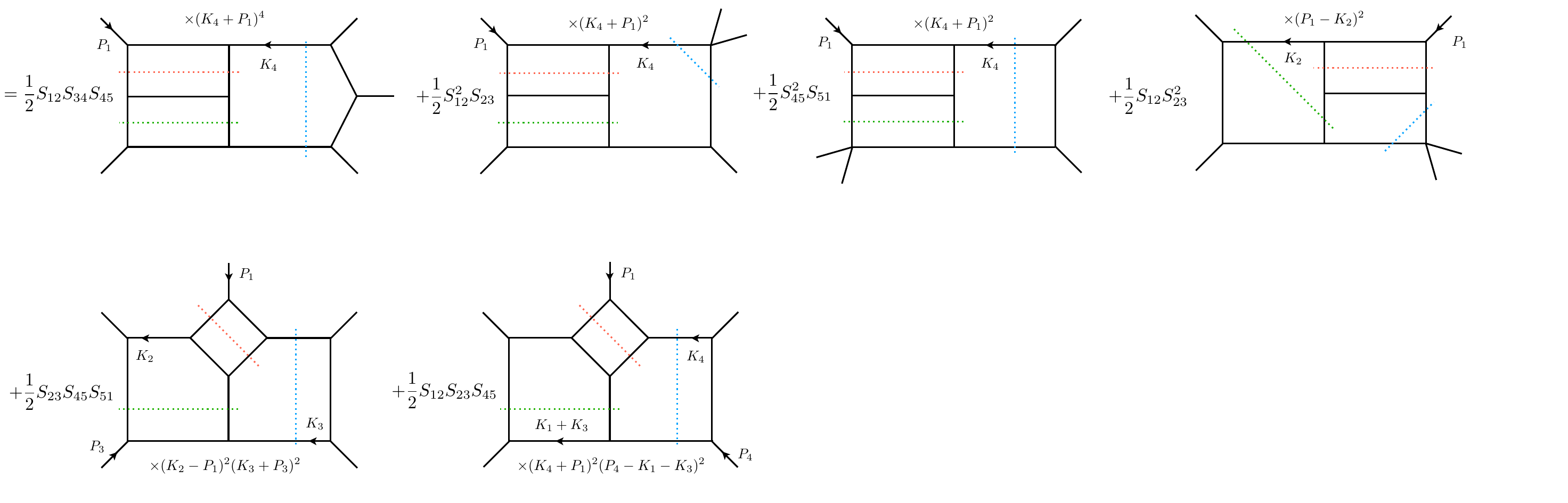}
}
\nonumber
\end{align}

\begin{figure}[t]
\begin{center}
\mbox{
\begin{picture}(0,285)(217,0)
\put(0,0){\insertfig{15}{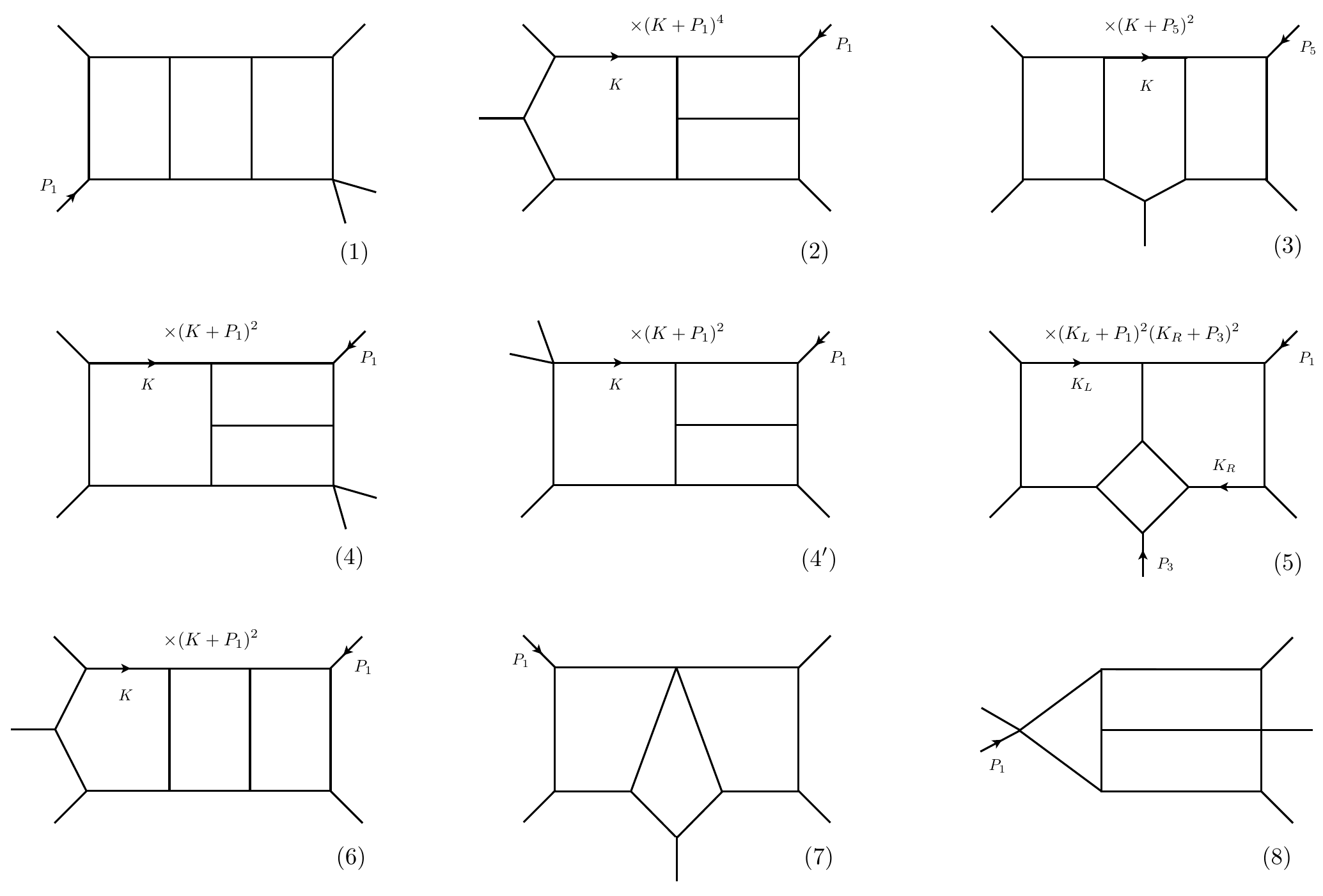}}
\end{picture}
}
\end{center}
\caption{\label{3LMIs} The set of master integrals defining the three-loop five-leg amplitude. Enumeration of legs always starts
from the top left corner of each graph and runs counterclockwise.}
\end{figure}

We can immediately identify nine six-dimensional master integrals in the basis. They are shown in Fig.~\ref{3LMIs} such that the 
amplitude reads
\begin{align}
\label{AssemblyEq}
\mathcal{A}^{(3)}_5
=
\mathcal{A}^{(0)}_5 
\sum_{\sigma_5 \cup \bar\sigma_5}
\Big[
C_{1} \mathcal{I}_1
+
\ft12 C_{2} \mathcal{I}_2
&
+
C_{3} \mathcal{I}_3
+
C_{4} \mathcal{I}_4
+
C_{4'} \mathcal{I}_{4'}
\\[-2mm]
&
+
C_{5} \mathcal{I}_5
+
\ft12 C_{6} \mathcal{I}_6
+
\ft12
C_{7} \mathcal{I}_7
+
\ft12 C_{8} \mathcal{I}_8
\Big]
\, . \nonumber
\end{align}
Here, the sum runs over the elements of the dihedral group $D_5$ formed by the five cyclic permutations $\sigma_5$ and their reflections
$\bar\sigma_5$
\begin{align}
\sigma_5
&= \{(12345), (23451), (34512), (45123), (51234) \}
\, , \\
\bar\sigma_5
&= \{(54321), (43215), (32154), (21543), (15432) \}
\, .
\end{align}
The accompanying fractions eliminate double-counting. The coefficients found from the six-dimensional cuts are determined
by the Mandelstam invariants and read
\begin{alignat}{6}
&C_1 &&= - \frac{1}{2} S_{51}^3 S_{45}
\, , \qquad
&&C_2 &&= - \frac{1}{2} S_{23} S_{34} S_{51}
\, , \qquad
&&C_3 &&= - \frac{1}{2} S_{12}^2 S_{23} S_{45}
\, , \nonumber\\
&C_4 &&= - \frac{1}{2} S_{12}^2 S_{23}
\, , \qquad
&&C_{4'} &&= - \frac{1}{2} S_{45} S_{51}^2
\, , \qquad
&&C_5 &&= - \frac{1}{2} S_{23} S_{34} S_{51}
\, , \label{eq:coefficients}\\
&C_6 &&= - \frac{1}{2} S_{23} S_{34} S_{51}^2
\, , \qquad
&&C_7 &&= \frac{1}{2} S_{12} S_{23} S_{34} S_{45}
\, , \qquad
&&C_8 &&= \frac{1}{2} S_{51} S_{12} S_{34}
\, . \nonumber
\end{alignat}
Notice that we adopted the nomenclature of Ref.\ \cite{Ambrosio:2013pba} for enumerating the integrals in the basis and added
explicitly the element $(4')$ since it arises in Ref.\ \cite{Ambrosio:2013pba} from the use of a `cyc'-operation introduced there,
while we merely use $D_5$ permutations. Our consideration confirms that the six-dimensional five-leg amplitude at three loops is
merely an uplift of all Lorentz-invariant products \cite{Spradlin:2008uu,Ambrosio:2013pba} from four to six dimensions.

\section{Dimensional reduction to special Coulomb branch}
\label{SectionDR}

The higher-dimensional vantage point employed in the previous sections provides an exhaustive description of the Coulomb branch of
$\mathcal{N} = 4$ sYM: the first four components of the six-dimensional momenta are the usual four-dimensional momenta of particles,
while the remaining two are interpreted as masses. As our main interest is in the off-shell amplitudes, i.e., when all particles propagating
in loops are massless, while the external asymptotic legs are massive, or off-shell. We need to dimensionally reduce the integral basis
of the $\mathcal{N} = (1,1)$ amplitude down to the special Coulomb branch of the four-dimensional ${\mathcal N} =4$ sYM.  To this end,
first, we pass to the dual variables $X$,
\begin{align}
P_i = X_{i} - X_{i + 1}
\, ,
\end{align}
which are decomposed in terms of their four- and out-of-four-dimensional components,
\begin{align}
X_i = (x_i, y_i)
\, .
\end{align}
Here and below it is tacitly implied that all indices are defined $\mbox{mod}(5)$, i.e., $6 = 1$, etc.

The $x$'s are the regions' momenta in four dimensions, such that the four-dimensional momenta are
\begin{align}
p_i = x_{i} - x_{i + 1}
\, .
\end{align}

The extra-dimensional $y$-vectors encode the mass/off-shellness effects. We parametrize them as
\begin{align}
y_i = m n_i
\, ,
\end{align}
with a mass parameter $m$. To descend onto the special Coulomb branch of the theory, we choose $n_i$'s to be light-like\footnote{This
can be accomplished by complexification as well.}
\begin{align}
n_i^2 = 0
\, ,
\end{align}
so that all internal lines in Feynman graphs are strictly massless. To induce identical external virtualities, we set the product of adjacent
vectors in the color-ordered amplitudes to be
\begin{align}
n_i \cdot n_{i + 1} = \ft12
\, .
\end{align}
The masslessness condition of the six-dimensional momenta $P_i^2 = 0$ then implies  the off-shell condition for the four-dimensional
momenta $p_i^2 = - y_{i,i+1}^2 = m^2$. The next-to-neighboring products are parametrized by a set of angles $\theta$ in the
extra-dimensional subspace ($j \neq i+1$)
\begin{align}
n_i \cdot n_j = \cos \theta_{ij}
\, .
\end{align}
However, they will not play any role in our consideration since they only emerge, upon dimensional reduction, in the six-dimensional
Mandelstam invariants down to four,
\begin{align}
\label{6Dto4D}
S_{i,i+1} = s_{i,i+1} + 2 m^2 \cos\theta_{i,i+2}
\, ,
\end{align} 
and will not produce a finite contribution in the limit $m \to 0$. Here $s_{i,i+1} \equiv (p_i + p_{i+1})^2$ are defined in the Lorentzian signature

After passing to the above dual variables, the momentum integrations, accompanied by a factor of the six-dimensional gauge coupling
$g_6$ and the number of colors $N_c$ (in the planar limit), are kept strictly in four dimensions,
\begin{align}
g_6^2 N_c \int \frac{d^6 K}{(2 \pi)^6} \stackrel{\rm dim.red.}{\to} g^2 \int \frac{d^4 k}{\pi^2}
\, .
\end{align}
Here, $g^2 = g_4^2 N_c/(4 \pi)^2$ is the four-dimensional 't Hooft coupling. All loop integrals on this branch,
$\mathcal{I}_\alpha \stackrel{\rm dim.red.}{\to} I_\alpha$, are finite since the external off-shellness provides
a natural infrared regulator.

\section{Three-loop amplitude}
Now we are ready to present the five-leg amplitude on the special Coulomb branch $\mathcal{A}_5 \stackrel{\rm dim.red.}{\to} A_5$,
making use of the results established in preceding sections. We will do it for the ratio function $M_5$ defined as
\begin{align}
A_5 = A^{(0)}_5 M_5
\, .
\end{align}
The dependence on quantum numbers of particles involved in scattering enters only through the tree amplitude $A^{(0)}_5$.
$M_5$ develops a series in t' Hooft coupling
\begin{align}
M_5 = 1 + g^2 M_5^{(1)} + g^4 M_5^{(2)} + g^6 M_5^{(3)} + \dots
\, .
\end{align}

For explicit representation of our results, it is convenient to absorb the Mandelstam prefactors \re{eq:coefficients} into the definition 
of master integrals such that they become dimensionless. Another advantage of doing this is that they become manifestly dual 
conformal invariant! They develop kinematic dependence on cross ratios only. One set of these is given by 
\begin{align}
\label{vs}
v_i \equiv
\frac{x_{i+1, i-1}^2 x_{i+2, i-2}^2}{x_{i+1, i-2}^2 x_{i+2, i-1}^2}
\, .
\end{align}
for $i = 1,\dots, 5$. They can be easily recast in terms of four-dimensional Mandelstam invariants $s_{i,i+1}$ and the off-shellness $m$
\begin{align}
v_i = 
\frac{m^2 s_{i-1,i}}{s_{i+1,i+2} s_{i+2,i+3}}
\, .
\end{align}
Yet another useful set of cross ratios
\begin{align}
u_i
\equiv
\frac{x_{i+1, i+2}^2 x_{i-2, i-1}^2}{x_{i+1, i-2}^2 x_{i+2, i-1}^2}
=
\frac{m^4}{s_{i+1,i+2} s_{i+2,i+3}}
\, ,
\end{align}
is related to the above via $u_i=v_{i-1}v_{i+1}$.

Finally, calculations of all master integrals were performed in the Euclidean kinematics such that $s_{i,i+1} < 0$ and $m^2 < 0$. 
Obviously, the conformal cross-ratios do not depend on these signs. However, we will incorporate additional minus signs explicitly 
in the definition of Lorentzian propagators, irreducible scalar products. According to this nomenclature, at the risk of being repetitive, 
our three-loop four-dimensional master integrals are given by the formula
\begin{align}
I_\alpha^{(3)}
=
\int \prod_{\ell = 1}^3 \frac{d^4 k_\ell}{i \pi^2} \mbox{num}_\alpha \prod_{p \in S_\alpha} D^{-a_p}_p
\, .
\end{align}
Here, $S_\alpha$ for each diagram encodes (inverse) massless propagators and irreducible scalar products, schematically
$D_p \equiv - (k_\ell + \dots)^2$, as well as their corresponding powers $a_p$ taking values in the set $\{1,0,-1,-2 \}$. 
These can be readily read off from the graphs shown in Fig.\ \ref{3LMIs}. The loop momentum-independent numerators are
\begin{align}
\mbox{num}_\alpha = 
\{
s_{51}^3 s_{45}
, 
-
s_{23} s_{34} s_{51}
, 
s_{12}^2 s_{23} s_{45}
,
&
-
s_{12}^2 s_{23}
,
-
s_{45} s_{51}^2
,
\\
&
-
s_{23} s_{34} s_{51}
,
s_{23} s_{34} s_{51}^2
,
s_{12} s_{23} s_{34} s_{45}
, 
-
s_{51} s_{12} s_{34}
\}
\, . \nonumber
\end{align}
Additional sporadic minus signs compensate for the Lorentzian nature of Mandelstam variables defined in Eq.\ \re{6Dto4D}.
The three-loop assembly equation becomes very simple
\begin{align}
M_5^{(3)} 
\label{AssemblyEq3L} 
=
\sum_{\sigma_5 \cup \bar\sigma_5}
\left[
- 
\ft12 I^{(3)}_1
-
\ft14 I^{(3)}_2
-
\ft12 I^{(3)}_3
-
\ft12 I^{(3)}_4
-
\ft12 I^{(3)}_{4'}
-
\ft12 I^{(3)}_5
-
\ft14 I^{(3)}_6
+
\ft14 I^{(3)}_7
+
\ft14 I^{(3)}_8
\right]
\, .
\end{align}
It has to be supplemented by similar equations at one and two loop orders, which were established in Ref.\ \cite{Belitsky:2025bgb}
and read
\begin{align}
\label{AssemblyEq1loop}
M^{(1)}_5
=
\sum_{\sigma_5 }
\ft12 {I}_1^{(1)}
\, , \qquad
M^{(2)}_5
=
\sum_{\sigma_5 }
\left[
{I}_1^{(2)} + \ft12 {I}_2^{(2)}
\right]
\, ,
\end{align}
in terms of the box ${I}_1^{(1)}$, double box ${I}_1^{(2)}$, and pentabox ${I}_2^{(2)}$ integrals. Technical details on the calculation of the 
three-loop integrals will be presented in a companion paper \cite{BBLOS2026}. For the reader's convenience, we provide their explicit  
expressions in the near-mass shell limit in the Appendix \ref{IntegralsAppendix} and in the accompanying ancillary Mathematica files.

Substituting our integral solutions into the three-loop assembly equation \re{AssemblyEq3L} and adding this three-loop correction to our 
previously established two-loop amplitude from Ref.\ \cite{Belitsky:2025bgb}, we immediately observe the exponentiation of the amplitude 
to this order in 't Hooft coupling, such that its logarithm, up to a kinematically independent constant, admits the following form
\begin{align}
\label{M5PTuBasis}
\log M_5
&
=
\left(
- \ft14 g^2 + \zeta_2 g^4  - \ft{27}{2} \zeta_4 g^6
\right)
\mathbb{L}_0^2 (u)
\\
&
+
\left(
- \ft12 g^2 + \zeta_2 g^4 - \ft{27}{2} \zeta_4 g^6
\right)
\mathbb{L}_{1}^2 (u)
+
\left(
\ft12 g^2 -  \zeta_2 g^4 + 11 \zeta_4 g^6
\right)
\mathbb{L}_{2}^2 (u)
+
d(g)
\, . \nonumber
\end{align}
Or, equivalently
\begin{align}\label{M5PTvBasis}
\log M_5
=
&
\left(  \zeta_2 g^4 - 16 \zeta_4 g^6 \right) \mathbb{L}^2_0 (v)
\\
+
&
\left( - g^2 + 2 \zeta_2 g^4 - \ft{59}{2} \zeta_4 g^6 \right)
\mathbb{L}^2_{1} (v)
+
\left(  \zeta_2 g^4 - \ft{37}{2}  \zeta_4 g^6 \right)
\mathbb{L}^2_{2} (v)
+
d(g)
\, , \nonumber
\end{align}
in terms of the logarithms of the $v$-cross ratios. Above, we introduced short-hand notations for linear combinations of double logarithms
of cross ratios\footnote{We will respectively call them the $u$- and $v$-bases.} $w = (u \ \mbox{or}\ v)$ involving coincident (0), nearest
neighbor (1), and next-to-nearest neighbor (2) values of their indices,
\begin{align}
\mathbb{L}^2_0 (w) = \sum_{i = 1}^5 \log^2 w_i
\, ,
\qquad
\mathbb{L}^2_{1} (w) = \sum_{i = 1}^5 \log w_i \log w_{i+1}
\, ,
\qquad
\mathbb{L}^2_{2} (w) = \sum_{i = 1}^5 \log w_i \log w_{i+2}
\, .
\end{align}
It is safe to assume that the following form holds for the all-order five-leg amplitude
\begin{align}
\log M_5
=
\gamma_{0} (g) ~\mathbb{L}^2_0 (u)
+
\gamma_{1}(g) ~\mathbb{L}^2_{1} (u)
+
\gamma_{2}(g) ~\mathbb{L}^2_{2} (u) + d(g)
\, ,
\end{align}
with $\gamma$'s and $d$ being functions of 't Hooft coupling only. The same expression in terms of the logarithms of the $v$-cross ratios will be given by:
\begin{align}
\log M_5
=&
\big( 2\gamma_{0} (g) + \gamma_{2} (g) \big) ~\mathbb{L}^2_0 (v)
+
\big( 3\gamma_{1} (g) + \gamma_{2} (g) \big) ~\mathbb{L}^2_{1} (v)
\nonumber\\
+&
\big( 2\gamma_{0} (g) + \gamma_{1} (g) + 2\gamma_{2} (g) \big) ~\mathbb{L}^2_{2} (v) + d(g)
\, .
\end{align}

Several comments are in order regarding these expressions.  First, the universality of infrared logarithms on the special Coulomb branch
should manifest itself with the emergence of the octagon anomalous dimension for the double logarithmic behavior in the particle off-shellness
$m$. Indeed, keeping track of $\log^2 m^2$ only, we do find
\begin{align}
\log M_5|_{\log^2 m^2} = - \ft{5}{4} \Gamma_{\rm oct} (g) \log^2 m^2
\, ,
\end{align}
where \cite{Coronado:2018cxj,Belitsky:2019fan}
\begin{align}
\Gamma_{\rm oct} (g) = 4 g^2 - 16 \zeta_2 g^4 + 256 \zeta_4 g^6 + \dots
\, .
\end{align}

In fact, we have an even stronger constraint on the infrared behavior of the amplitude, namely, all singular terms are condensed in the
product of Sudakov form factors on the Coulomb branch of the theory:
\begin{align}
\label{AmplFactorOffShellgeneralForm}
\log M_5 = \ft{1}{2}\sum_{i=1}^5
\log \mathcal{F}_{2}^{\rm off} \left(s_{i,i+1}/m^2; g\right) + f_5\big(\{s_i\},g\big) + O(m^2),
\end{align}
where
\begin{align}
\label{FFFactorOffShell}
\log \mathcal{F}_{2}^{\rm off} \left(s_{i,i+1}/m^2; g\right)
=
-
\ft{1}{2}\Gamma_{\rm oct} (g) \log^2 \left(s_{i,i+1}/m^2 \right)
-
D(g) + O(m^2).
\end{align}
Presently, the explicit form of $D(g)$ is not important for us. It can be found in Ref.\ \cite{Belitsky:2022itf}. In full agreement with the general
expectations regarding the structure of IR divergences on the Coulomb branch, we confirm that (\ref{M5PTuBasis}) and (\ref{M5PTvBasis})
also satisfy (\ref{AmplFactorOffShellgeneralForm}) constraint. This also implies that the $\gamma$-functions are not independent and should
be related to each other via the {\rm sum rule} to all orders of perturbation theory
\begin{align}
\label{SumRule}
\gamma_{0} (g) + \gamma_{1} (g) + \gamma_{2} (g)
=
- \ft{1}{16}
\Gamma_{\rm oct} (g)
\, .
\end{align}

The exponentiated structure of the amplitude in either of the bases, (\ref{M5PTuBasis}) and (\ref{M5PTvBasis}), resembles expressions
deduced for correlation functions in a light cone limit \cite{Bercini:2020msp} or the origin limit of six-gluon amplitudes in the purely
massless case \cite{Basso:2020xts}. It is, therefore, tempting to cast $\gamma$-functions as linear combinations of the so-called
{\sl tilted} anomalous dimensions $\Gamma_{\alpha}(g)$. These are defined as solutions\footnote{For the reader's convenience, we give
their explicit perturbative expansion in Appendix \ref{TiltedAppendix}.} of the deformed flux-tube integral equation with a deformation
parameter $\alpha$ \cite{Basso:2020xts}.  For example, for different values of the {\sl tilt angle} $\alpha$, we obtain either the cusp
$\Gamma_{\rm cusp} = \Gamma_{\pi/4}$ or the octagon $\Gamma_{\rm oct} = \Gamma_{0}$ anomalous dimensions.

In the $u$-basis, the coefficient in front of $\mathbb{L}^2_{2} (u)$ matches $\ft18 \Gamma_{\rm cusp}(g)$ in the first three perturbative
orders, suggesting that
\begin{align}
\label{Rel1}
\gamma_{2} (g) = \ft18 \Gamma_{\rm cusp}(g)
\, .
\end{align}
Assuming that this equality will hold at higher loops, and combining this observation with the sum rule \re{SumRule}, we can write the
following all-loop conjectures for the remaining two functions $\gamma_{0}(g)$ and $\gamma_{1}(g)$ as follows
\begin{alignat}{2}
\label{Rel23}
&
\gamma_{0} (g)
=
&&
- \ft{1}{48} \Gamma_{\rm oct} (g)- \ft{1}{8} \Gamma_{\rm cusp} (g)+ \ft{1}{12} \Gamma_{\rm hex} (g)
\, ,
\nonumber\\
&
\gamma_{1} (g)
=
&&
- \ft{1}{24} \Gamma_{\rm oct} (g) - \ft{1}{12} \Gamma_{\rm hex} (g)
\, ,
\end{alignat}
where $\Gamma_{\rm hex}=\Gamma_{\pi/3}$. However, such conjectures are not unique, since they can be freely altered by allowing for
an arbitrary combination involving $\Gamma_{\rm cusp}$ and $\Gamma_{\rm oct}$ in addition to a function proportional to $\Gamma_{\alpha}$
for a fixed value of $\alpha$. Our three-loop analysis does not rule out such a freedom. This function can even be chosen as the one-loop
exact $\Gamma_{\pi/2}=4 g^2$ together with just $\Gamma_{\rm cusp}$ and $\Gamma_{\rm oct}$, which yields a closed-form expression for
the $\gamma$-functions. This is not a consistent conjecture, however, due to the strong-coupling constraint on $\Gamma_{\alpha}$. The main
problem with $\Gamma_{\pi/2}$ is that it cannot appear at strong coupling since its $g^2$-behavior would contradict the classical string picture
stemming from the AdS/CFT correspondence for scattering amplitudes \cite{Alday:2007hr,Alday:2007he}. Notice that the strong-coupling
solutions to flux-tube equations yield the following strong coupling behavior for the tilted dimension \cite{Basso:2020xts}
\begin{align}
\Gamma_\alpha (g)|_{g \to \infty} = \frac{8 \alpha}{\pi \sin(2 \alpha)} g + O (g^0)
\, ,
\end{align}
such that $\Gamma_{\pi/2}$ diverges. Unfortunately, with the perturbative data available and the all-loop constraint (\ref{SumRule}),
we could not unambiguously and uniquely fix $\gamma$'s in terms of tilted anomalous dimensions. So additional information is required.
This is what we turn to next.

There is another source of information regarding the all-loop structure of the $M_5$-amplitude. As was suggested in Refs.\
\cite{Caron-Huot:2021usw,Bercini:2024pya,Belitsky:2025bgb}, there is a conjectured duality between correlation functions of
infinitely heavy half-BPS operators and scattering amplitudes on the Coulomb branch. For the five-point correlation function/amplitude,
the relation looks like
\begin{align}
\label{DualityGM}
\log\mathbb{D}_0 = \log M_5 + O(m^2)
\, .
\end{align}
where $\mathbb{D}_0$ is defined from the pair-wise null limit
\begin{align}
\lim_{x^2_{i,i+1} \to 0} \lim_{K \to \infty} G_{5, K}/G_{5, K}^{\rm tree} = \mathbb{D}_0^2 (u_1,\dots, u_5; g)
\, ,
\end{align}
of the correlator\footnote{Here, $X,Y,Z$ and $\bar{X},\bar{Y},\bar{Z}$ are the complex scalars and their complex conjugates in $\mathcal{N} = 4$ sYM.}
$G_{5, K}$,
\begin{equation}
\label{Korr51}
G_{5,K}
=
\big\langle{\rm tr}
X^{2K}(x_1) \ {\rm tr}\bar{X}^K\bar{Y}^K (x_2) \ {\rm tr}Y^K\bar{Z}^K (x_3) \ {\rm tr}Z^{2K}(x_4) \ {\rm tr}\bar{Z}^K\bar{X}^K(x_5)
\big\rangle
\, .
\end{equation}
This duality was tested up to two loops \cite{Bercini:2024pya,Belitsky:2025bgb}, and our three-loop computation can be considered as a
prediction for the correlation function side of this duality. One can hope that integrability-based method of hexagonalization
\cite{Basso:2015zoa,Fleury:2016ykk,Fleury:2017eph} will eventually allow one to obtain information regarding the behavior of $M_5$ at
higher orders of perturbative series.

In conclusion, let us also make the following general observations. First, while individual master integrals contain odd values of the Riemann
zeta functions $\zeta_{2k+1}$, they {\sl all} cancel in the amplitude. A similar phenomenon was observed in many other infrared-sensitive
observables on the special Coulomb branch
\cite{Caron-Huot:2021usw,Bork:2022vat,Belitsky:2022itf,Belitsky:2023ssv,Belitsky:2024agy,Belitsky:2024dcf}. While for
leading infrared double logarithms this is warranted by their universality, it also persists for subleading ones as well as for the finite
contributions. It is unclear whether it will persist at four loops and higher. It currently remains a mystery.

Second, and this is the most puzzling feature of our results, the anticipated rigidity of the kinematic structure of the all-order conjecture for
the five-leg amplitude put out in Ref.\ \cite{Belitsky:2025bgb} is lifted. We observe that the logarithms of cross ratios involving coincident,
nearest neighbor, and next-to-nearest neighbor values of their indices are accompanied by different functions of the 't Hooft coupling.
While in itself it is not a problem, what is troublesome is, as was stated before, we failed to find nice combinations of tilted
anomalous dimensions $\Gamma_\alpha$ that naturally incorporate the octagon anomalous dimension $\Gamma_{\rm oct}$, free from odd
values of the Riemann zeta function at four loops and higher, and do not include $\Gamma_{\pi/2} = 4 g^2$, which is one-loop exact.
If one forfeits the appearance of $\Gamma_{\rm oct}$ in the all-order expression for the five-leg amplitude, one can play numerology games
and devise multiple expressions that match the above explicit three-loop calculation. A particular one that stands out is the following
\begin{align}
\log M_5 = - \ft18 \Gamma_{\pi/12}  \left[\mathbb{L}^2_{1} + \ft{1}{\sqrt{3}} \mathbb{L}^2_0 \right]
-
\ft18 \Gamma_{5\pi/12} \left[\mathbb{L}^2_{1} - \ft{1}{\sqrt{3}} \mathbb{L}^2_0 \right]
+
\ft1{4\sqrt{5}} [ \Gamma_{3\pi/10} - \Gamma_{\pi/10} ] \mathbb{L}^2_{2}
\, .
\end{align}
It is distinguished by the cancellation of odd values of the Riemann zeta function at four loops; however, the four-loop coefficient
of the double infrared log is no longer given by the octagon anomalous dimension, though the difference is very small $\ft{2141}{630}
- \ft{1088}{315} = -\ft{1}{18}$. Going beyond $g^8$-order, the presence of $\zeta_{2n+1}$ becomes unavoidable.

Speculations in the previous paragraph are, however, very unnatural in light of the expected universality of infrared logarithms. So we have
to insist on the presence of $\Gamma_{\rm oct}$ as the coefficient of $\log^2 m^2$. Then various combinations of $\Gamma_{\alpha}$ present
themselves; however, they are all plagued by odd zeta's already at four loops. In light of this, there is no preferred choice. The most aesthetically
pleasing combination was already quoted in Eqs.\ \re{Rel1}-\re{Rel23}. Therefore, without further input from an explicit perturbative calculation
at $O(g^8)$, we cannot offer a conjecture for the all-order form of the five-leg amplitude.

\section{Conclusion}

In this work, we constructed the three-loop integral basis for the five-leg amplitude on the Coulomb branch of $\mathcal{N} = 4$ sYM.
We relied on the six-dimensional formulation of the theory, which enjoys an unconstrained spinor-helicity formalism and unitarity
cut-sewing technique. The six-dimensional basis of integrals was then dimensionally reduced to four dimensions. The extra-dimensional
components of momenta were interpreted as masses. The integrals and their accompanying coefficients are a naive dimensional uplift
of the previously known integrals in purely massless kinematics.

All of the three-loop integrals were computed by us in a companion paper in the near-massless limit of external legs and massless internal
propagators. This kinematics mimics the off-shell situation of massless scattering, which is the main objective of our study. The amplitude
inherits the exact dual conformal symmetry of the four-dimensional scattering. As compared to massless scattering, all expressions are
finite with the infrared regulator being given by the external off-shellness $m$. For the five-leg amplitude at hand, all independent dual-conformal
cross ratios are small in the limit $m \to 0$. As a result, all transcendental functions of cross ratios degenerate into logarithms. Thus, we
found a very concise representation for it for the three-loop correction.

One of the goals of the present study was to verify whether the kinematical dependence of the all-loop conjecture put out in Ref.\
\cite{Belitsky:2025bgb} is rigid against higher-loop effects. We found that it is not! Namely, three possible kinematical structures involving
logarithms of coincident, nearest neighbor, and next-to-nearest neighbor cross-ratios are accompanied by independent functions of the
coupling. The infrared double logarithms are indeed driven by the octagon anomalous dimension, in agreement with the universality of
the former in the off-shell setup. Recall that for the on-shell case, the relevant anomalous dimension is the cusp
\cite{Sterman:2002qn,Aybat:2006mz,Dixon:2008gr,Becher:2009qa,Anastasiou:2003kj,Bern:2005iz}. However, our findings
at three loops do not apparently allow for a natural representation
of the found functions of 't Hooft coupling in terms of tilted anomalous dimensions that correctly incorporate expected transcendental numbers
appearing in front of infrared logarithms and have a well-defined behavior at strong coupling in accord with the AdS/CFT dictionary. This begs
for a four-loop extension of our results. In light of the recent progress in multiloop calculations, which we report on in the companion paper,
this could just be feasible.

\begin{acknowledgments}

We are grateful to Erik Panzer for his valuable advice regarding the application of {\tt HyperInt}. R.L. is
would like to thank Bukhard Eden and Paul Heslop for explaining the action of the `cyc' operator from Ref.\ \cite{Ambrosio:2013pba}.
The work of V.S. was supported by the Moscow Center for Fundamental and Applied Mathematics of Lomonosov Moscow State 
University under Agreement No.\ 075-15-2025-345. L.B.\ was supported by the Foundation for the Advancement of Theoretical 
Physics and Mathematics “BASIS”.

\end{acknowledgments}

\appendix

\section{Loop integrals}
\label{IntegralsAppendix}

\begin{figure}[t]
\begin{center}
\mbox{
\begin{picture}(0,90)(210,0)
\put(0,0){\insertfig{14}{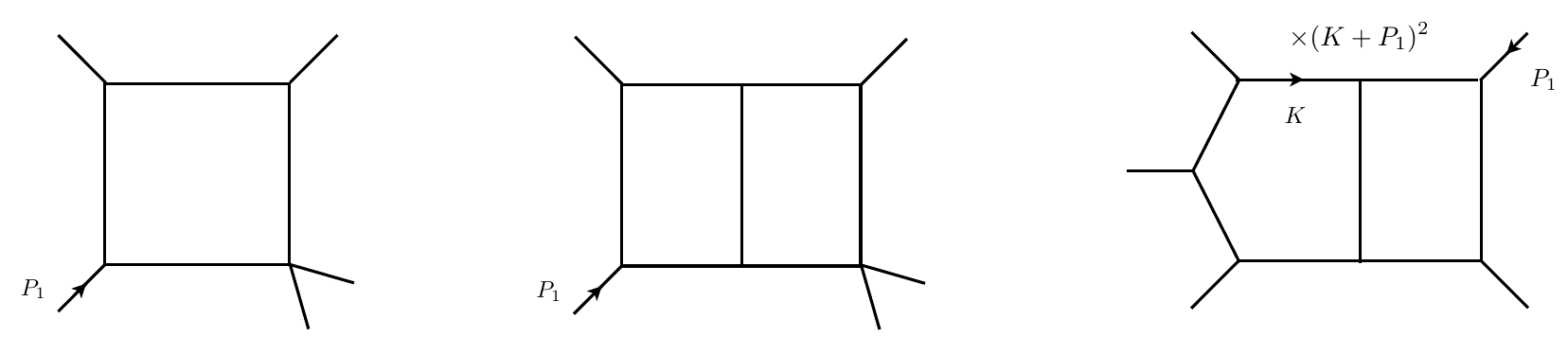}}
\end{picture}
}
\end{center}
\caption{\label{1-2LMIs} Master integrals at one (Box) and two loops (Dbox and Pentabox). Their analytical representation
follows the same nomenclature as in the main text. The additional loop momentum-independent numerators are $\mbox{num} 
= \{ s_{51} s_{45}, - s_{51}^2 s_{45}, - s_{23} s_{34} s_{51} \}$, respectively. Again, with extra signs introduced to compensate 
for the Lorentzian nature of the Mandelstam invariants.}
\end{figure}

Here, we present explicit solutions for one-, two-, and three-loop integrals that enter the assembly equations 
\re{AssemblyEq3L}-\re{AssemblyEq1loop}. To make explicit expressions as concise as possible, we introduce 
the notation 
\begin{align}
L_i \equiv \log v_i \, ,
\end{align}
everywhere below.  The first two orders are shown in Fig.\ \ref{1-2LMIs} and read \cite{Belitsky:2025bgb}
\begin{align}
I_1^{(1)}
&
=
L_{3} L_{4} + L_{4}L_{5} + 2 \zeta_2
\, , \\
I_1^{(2)}
&
=
\ft{1}{4} L_{4}^{2}(L_{3} + L_{5})^{2} + \ft{1}{2} (L_{3}^{2} + 4L_{3}L_{4} 
+ 
L_{4}^{2} + 2L_{3}L_{5} + 4L_{4}L_{5} + L_{5}^{2})\zeta_{2} + \ft{21}{2} \zeta_{4}
\, , \\
I_2^{(2)}
&
=
\ft{1}{2} L_{1}(L_{2}^{2}L_{3} + 2L_{2}L_{3}L_{4} + 2L_{3}L_{4}L_{5} + L_{4}L_{5}^{2}) 
\nonumber\\
&
+ 
\ft{1}{2} (4L_{1}L_{2} - L_{2}^{2} + 2L_{1}L_{3} + 2L_{1}L_{4} - 2L_{2}L_{4} + 4L_{3}L_{4} + 4L_{1}L_{5} - 2L_{3}L_{5} - L_{5}^{2})\zeta_{2} 
\nonumber\\
&
+
(L_{3} + L_{4} - 2L_{1})\zeta_{3} + 5\zeta_{4}
\, , 
\end{align}
while $O (g^6)$ is \cite{BBLOS2026}
\begin{align}
I^{(3)}_1 =
\,
&\ft{1}{36} L_3^3 (L_2 + L_4)^3
\nonumber\\
+
\,
&\ft{1}{6} L_3 (L_2 + L_4) (L_2^2 + 3 L_2 L_3 + L_3^2 + 2 L_2 L_4 + 3 L_3 L_4 + L_4^2) \zeta_2
\nonumber\\
+
\,
& \ft{7}{2} (L_2^2 + 3 L_2 L_3 + L_3^2 + 2 L_2 L_4 + 3 L_3 L_4 + L_4^2) \zeta_4
+
\ft{155}{4} \zeta_6
\, , \\[2mm]
I^{(3)}_2 =
\,
&\ft{1}{12}
(L_1 L_2^3 L_3^2 + 3 L_1 L_2^2 L_3^2 L_4 + 3 L_1 L_2 L_3^2 L_4^2 + 3 L_1 L_3^2 L_4^2 L_5 + 3 L_1 L_3 L_4^2 L_5^2 + L_1 L_4^2 L_5^3)
\nonumber\\
+
\,
&\ft{1}{6} (L_1 L_2^3 + 6 L_1 L_2^2 L_3 - L_2^3 L_3 + 3 L_1 L_2 L_3^2 + 3 L_1 L_2^2 L_4 + 12 L_1 L_2 L_3 L_4 - 3 L_2^2 L_3 L_4
\nonumber\\
&\quad + 3 L_1 L_3^2 L_4 + 3 L_1 L_2 L_4^2 + 3 L_1 L_3 L_4^2 - 3 L_2 L_3 L_4^2 + 3 L_3^2 L_4^2 + 3 L_1 L_3^2 L_5
\nonumber\\
&\quad + 12 L_1 L_3 L_4 L_5 - 3 L_3^2 L_4 L_5 + 3 L_1 L_4^2 L_5 + 3 L_1 L_3 L_5^2 + 6 L_1 L_4 L_5^2 - 3 L_3 L_4 L_5^2
\nonumber\\
&\quad + L_1 L_5^3 - L_4 L_5^3) \zeta_2
\nonumber\\
+
\,
&\ft{1}{6} (-L_2^3 - 3 L_1 L_3^2 - 3 L_2^2 L_4 - 12 L_1 L_3 L_4 + 3 L_3^2 L_4 - 3 L_1 L_4^2 - 3 L_2 L_4^2 + 3 L_3 L_4^2
\nonumber\\
&\quad - 3 L_3^2 L_5 - 3 L_3 L_5^2 - L_5^3) \zeta_3
-
(4 L_1 + L_2 - 5 L_3 - 5 L_4 + L_5) \zeta_2 \zeta_3
\nonumber\\
+
\,
&\ft{1}{4} (42 L_1 L_2 - 10 L_2^2 + 35 L_1 L_3 - 10 L_2 L_3 + 7 L_3^2 + 35 L_1 L_4 - 20 L_2 L_4 + 20 L_3 L_4
\nonumber\\
&\quad + 7 L_4^2 + 42 L_1 L_5 - 20 L_3 L_5 - 10 L_4 L_5 - 10 L_5^2) \zeta_4
\nonumber\\
-
\,
&4 L_1 \zeta_5 + 2 \zeta_3^2 + \ft{77}{8} \zeta_6
\, , \\[2mm]
I^{(3)}_3
=
\,
&\ft{1}{12} (2 L_1^3 L_2 + 3 L_1^2 L_2 L_3 + 3 L_1^2 L_2 L_4 + 6 L_1 L_2 L_3 L_4 + 3 L_2 L_3 L_4^2 + L_3 L_4^3) L_5^2
\nonumber\\
+
\,
&\ft{1}{6} (2 L_1^3 L_2 + 3 L_1^2 L_2 L_3 + 3 L_1^2 L_2 L_4 + 6 L_1 L_2 L_3 L_4 + 3 L_2 L_3 L_4^2 + L_3 L_4^3 - 2 L_1^3 L_5
\nonumber\\
&\quad + 6 L_1^2 L_2 L_5 - 3 L_1^2 L_3 L_5 + 12 L_1 L_2 L_3 L_5 - 3 L_1^2 L_4 L_5 - 6 L_1 L_3 L_4 L_5 + 12 L_2 L_3 L_4 L_5
\nonumber\\
&\quad - 3 L_2 L_4^2 L_5 - L_4^3 L_5 + 3 L_1^2 L_5^2 + 3 L_1 L_2 L_5^2 + 3 L_2 L_3 L_5^2 + 6 L_1 L_4 L_5^2
\nonumber\\
&\quad + 3 L_2 L_4 L_5^2 + 3 L_3 L_4 L_5^2 + 3 L_4^2 L_5^2) \zeta_2
\nonumber\\
+
\,
&\ft{1}{6} (-2 L_1^3 - 3 L_1^2 L_3 - 3 L_1^2 L_4 - 6 L_1 L_3 L_4 + 3 L_2 L_4^2 + L_4^3 + 6 L_2 L_4 L_5
\nonumber\\
&\quad + 6 L_3 L_4 L_5 + 3 L_2 L_5^2 - 6 L_4 L_5^2) \zeta_3
-
(L_1 - 3 L_2 - L_3 + 4 L_4 + 3 L_5) \zeta_2 \zeta_3
\nonumber\\
+
\,
&\ft{1}{4} (10 L_1 L_2 - 20 L_1 L_3 + 42 L_2 L_3 + 20 L_1 L_4 - 11 L_2 L_4 + L_3 L_4 + 10 L_4^2
\nonumber\\
&\quad + 30 L_1 L_5 + 6 L_2 L_5 + 4 L_3 L_5 + 20 L_4 L_5 + 10 L_5^2) \zeta_4
\nonumber\\
+
\,
&2 (L_2 + L_3) \zeta_5 - 3 \zeta_3^2 + 28 \zeta_6
\, , \\[2mm]
I^{(3)}_{4'}
=
\,
&\ft{1}{36} L_3^3 (L_2 + L_4)^3
+
\ft{1}{6} L_3 (L_2 + L_4) (L_2^2 + 3 L_2 L_3 + L_3^2 + 2 L_2 L_4 + 3 L_3 L_4 + L_4^2) \zeta_2
\nonumber\\
+
\,
&\ft{7}{2} (L_2^2 + 3 L_2 L_3 + L_3^2 + 2 L_2 L_4 + 3 L_3 L_4 + L_4^2) \zeta_4 + \ft{155}{4} \zeta_6
\, , \\[2mm]
I^{(3)}_5
=
\,
&\ft{1}{12} L_{2}
(
3 L_{1} L_{2} L_{3}^2 L_{4} + L_{2} L_{3}^3 L_{4} + L_{1}^3 L_{2} L_{5} + 3 L_{1}^2 L_{2} L_{4} L_{5}
+
6 L_{1} L_{2} L_{3} L_{4} L_{5} + L_{1}^3 L_{5}^2
\nonumber\\
&\quad + 3 L_{1}^2 L_{4} L_{5}^2 + 6 L_{1} L_{3} L_{4} L_{5}^2
)
\nonumber\\
+
&
\,
\ft{1}{6}
(
3 L_{1}^2 L_{2}^2 + 6 L_{1} L_{2}^2 L_{3} - 3 L_{1} L_{2} L_{3}^2 - L_{2} L_{3}^3 + 3 L_{1} L_{2}^2 L_{4} + 6 L_{1} L_{2} L_{3} L_{4}
+ 3 L_{2}^2 L_{3} L_{4}
\nonumber\\
&\quad
+ 3 L_{1} L_{3}^2 L_{4} + 6 L_{2} L_{3}^2 L_{4} + L_{3}^3 L_{4}
+ 6 L_{1}^2 L_{2} L_{5} + 3 L_{1} L_{2}^2 L_{5} - 3 L_{2}^2 L_{3} L_{5} + 12 L_{1} L_{2} L_{4} L_{5}
\nonumber\\
&\quad
+
12 L_{2} L_{3} L_{4} L_{5} + 3 L_{1} L_{2} L_{5}^2 - 3 L_{1} L_{3} L_{5}^2 - 3 L_{2} L_{3} L_{5}^2 + 6 L_{1} L_{4} L_{5}^2 + 3 L_{3} L_{4} L_{5}^2
) \zeta_{2}
\nonumber\\
+
\,
&
\ft{1}{6} (
- 6 L_{1} L_{2}^2 + 6 L_{1} L_{2} L_{3} + 3 L_{2}^2 L_{3} - 3 L_{1} L_{3}^2 - L_{3}^3 + 6 L_{1} L_{2} L_{4} + 6 L_{2}^2 L_{4}
- 6 L_{1} L_{3} L_{4}
\nonumber\\
&\quad
- 12 L_{2} L_{3} L_{4} + 3 L_{2}^2 L_{5}
+ 6 L_{1} L_{3} L_{5} - 6 L_{3} L_{4} L_{5} + 3 L_{1} L_{5}^2 + 3 L_{2} L_{5}^2
\nonumber\\
&\quad
+ 6 L_{3} L_{5}^2 - 3 L_{4} L_{5}^2
)
\zeta_{3}
-
(4 L_{1} + 7 L_{2} - 7 L_{3} - 2 L_{4} + 4 L_{5})
\zeta_{2} \zeta_{3}
\nonumber\\
+
\,
& \ft{1}{4}
(
41 L_{1} L_{2} + 3 L_{2}^2 + 12 L_{1} L_{3} + 20 L_{2} L_{3} - 10 L_{3}^2 + 9 L_{1} L_{4} + L_{2} L_{4} + 9 L_{3} L_{4} + 9 L_{1} L_{5}
\nonumber\\
&\quad
+ 28 L_{2} L_{5} - 22 L_{3} L_{5} + 31 L_{4} L_{5} - 4 L_{5}^2
)
\zeta_{4}
-
2 \left(3 L_{1} + L_{2} + 2 L_{3} - 2 L_{5}\right) \zeta_{5}
\nonumber\\
+\,
&
c_1 \zeta_6 + c_2 \zeta_3^2
\, , \\[2mm]
I^{(3)}_6
=
\,
&\ft{1}{12} L_1 (L_2^3 L_3^2 + 3 L_2^2 L_3^2 L_4 + 3 L_2 L_3^2 L_4^2 + 3 L_3^2 L_4^2 L_5 + 3 L_3 L_4^2 L_5^2 + L_4^2 L_5^3)
\nonumber\\
+
\,
&\ft{1}{6} (L_1 L_2^3 + 6 L_1 L_2^2 L_3 - L_2^3 L_3 + 3 L_1 L_2 L_3^2 + 3 L_1 L_2^2 L_4 + 12 L_1 L_2 L_3 L_4 - 3 L_2^2 L_3 L_4
\nonumber\\
&\quad + 3 L_1 L_3^2 L_4 + 3 L_1 L_2 L_4^2 + 3 L_1 L_3 L_4^2 - 3 L_2 L_3 L_4^2 + 3 L_3^2 L_4^2 + 3 L_1 L_3^2 L_5
\nonumber\\
&\quad + 12 L_1 L_3 L_4 L_5 - 3 L_3^2 L_4 L_5 + 3 L_1 L_4^2 L_5 + 3 L_1 L_3 L_5^2 + 6 L_1 L_4 L_5^2 - 3 L_3 L_4 L_5^2
\nonumber\\
&\quad + L_1 L_5^3 - L_4 L_5^3) \zeta_2
\nonumber\\
+
\,
&\ft{1}{6} (-L_2^3 - 3 L_1 L_3^2 - 3 L_2^2 L_4 - 12 L_1 L_3 L_4 + 3 L_3^2 L_4 - 3 L_1 L_4^2 - 3 L_2 L_4^2 + 3 L_3 L_4^2
\nonumber\\
&\quad - 3 L_3^2 L_5 - 3 L_3 L_5^2 - L_5^3) \zeta_3
-
(4 L_1 + L_2 - 5 L_3 - 5 L_4 + L_5) \zeta_2 \zeta_3
\nonumber\\
+
\,
&\ft{1}{4} (42 L_1 L_2 - 10 L_2^2 + 35 L_1 L_3 - 10 L_2 L_3 + 7 L_3^2 + 35 L_1 L_4 - 20 L_2 L_4 + 20 L_3 L_4
\nonumber\\
&\quad + 7 L_4^2 + 42 L_1 L_5 - 20 L_3 L_5 - 10 L_4 L_5 - 10 L_5^2) \zeta_4
\nonumber\\
-
\,
&4 L_1 \zeta_5 + 2 \zeta_3^2 + \ft{77}{8} \zeta_6
\, , \\[2mm]
I^{(3)}_7
=
\,
&\ft{1}{12} L_1 L_2  L_5 (L_2 + L_5) (2 L_1^2 + 3 L_1 L_3 + 3 L_1 L_4 + 6 L_3 L_4)
\nonumber\\
+
\,
&\ft{1}{2} (L_1^2 L_2^2 + 2 L_1 L_2^2 L_3 + L_2^2 L_3 L_4 + 4 L_1^2 L_2 L_5 + L_1 L_2^2 L_5 + 4 L_1 L_2 L_3 L_5 + 4 L_1 L_2 L_4 L_5
\nonumber\\
&\quad + 4 L_2 L_3 L_4 L_5 + L_1^2 L_5^2 + L_1 L_2 L_5^2 + 2 L_1 L_4 L_5^2 + L_3 L_4 L_5^2) \zeta_2
\nonumber\\
+
\,
&\ft{1}{6} (-4 L_1^3 - 6 L_1^2 L_3 - 3 L_2^2 L_3 - 6 L_1^2 L_4 + 3 L_2^2 L_4 - 12 L_1 L_3 L_4 - 6 L_2 L_3 L_4 + 6 L_2^2 L_5
\nonumber\\
&\quad - 6 L_3 L_4 L_5 + 6 L_2 L_5^2 + 3 L_3 L_5^2 - 3 L_4 L_5^2) \zeta_3 -(2 L_1 + 3 L_2 +3 L_5) \zeta_2 \zeta_3
\nonumber\\
+
\,
&\ft{1}{4} (40 L_1 L_2 + 3 L_2^2 + 31 L_2 L_3 - L_2 L_4 + 40 L_1 L_5 + 12 L_2 L_5 - L_3 L_5 + 31 L_4 L_5 + 3 L_5^2) \zeta_4
\nonumber\\
+
\,
&
2 (L_2 + L_5) \zeta_5  - 4 \zeta_3^2 + 28 \zeta_6
\, , \\[2mm]
I^{(3)}_8
=
\,
&
-
\ft{1}{3} (L_2 + L_5)^3 \zeta_3 - 2(L_2 + L_5) \zeta_2 \zeta_3 - 4\zeta_3^2 + 3(L_2 + L_5)^2 \zeta_4 - 10(L_2 + L_5) \zeta_5 + \ft{41}{2} \zeta_6
\, .
\end{align}
It might be a bit surprising to the reader that no transcendental functions, like Goncharov polylogarithms, appear in the results.
While this fact appears like a cancellation miracle \cite{Belitsky:2025sin} in approaches based on the conventional Method of
Regions \cite{Beneke:1997zp}, it becomes obvious in its dual conformal formulation of Ref.\ \cite{Bork:2025ztu}. Also, let us point
out that we failed to fix additive transcendental constants in the expression for $I^{(3)}_5$ for two contributing regions $99$ and
$163$, i.e.,  since {\tt Hyperint} \cite{Panzer:2014gra}, that was used in the analysis, could not finish the job in a sensible time frame,
with a time cut-off of 744 CPU hours per core on a station with 100GB RAM. Of course, they are given by a linear combination of
transcendental numbers $\zeta_6$ and $\zeta_3^3$. One could fix the accompanying rational coefficients $c_{1,2}$ from numerical
evaluations and the application of PSLQ; however, we were unable to achieve the required accuracy for this. Relying on the anticipated 
cancellation of Riemann zetas with off values of the argument, we can, in fact, fix $c_2 = -3$. Finally, we did not display $I^{(3)}_{4}$ 
explicitly since it contributes identically to the amplitude as its sibling $I^{(3)}_{4'}$.

\section{Tilted anomalous dimension}
\label{TiltedAppendix}

In the main text, we relied on the known perturbative series of the tilted anomalous dimension \cite{Basso:2020xts}. The latter
arises as a solution to flux-tube equations 
\begin{align}
\label{GCuspSolMod}
\Gamma_{\alpha}(g)
=
4g^2\left[\frac{1}{1+\mathbb{K}(\alpha)}\right]_{11}
\, ,
\end{align}
with the kernel
\begin{align}
\label{GCuspSolAlpha}
\mathbb{K}(\alpha)=2\cos(\alpha)
\begin{pmatrix}
\cos(\alpha)\mathbb{K}_{\circ\circ} & \sin(\alpha) \mathbb{K}_{\circ\bullet} \\
\sin(\alpha) \mathbb{K}_{\bullet\circ} & \cos(\alpha)\mathbb{K}_{\bullet\bullet} 
\end{pmatrix},
\end{align}
where the odd ($\circ$)/even ($\bullet$) elements correspond to even/odd elements of the matrix $(\mathbb{K})_{nm}$
\begin{eqnarray}
\label{KMatrixElements}
(\mathbb{K})_{nm}=2m(-1)^{m(n+1)}\int^{\infty}_0 \frac{dt}{t}\frac{J_n(2gt)J_m(2gt)}{e^t-1}.
\, .
\end{eqnarray}
Perturbative expansion is straightforward and yields
\begin{align}
\Gamma_\alpha (g)&= 4g^2
\\
& - 16 g^4 \cos^2 \alpha \, \zeta_2
\nonumber\\
& + 32 g^6 \cos^2\alpha \, (3 + 5\cos^2\alpha) \zeta_4
\nonumber\\
& - 128 g^8 \cos^2\alpha \left[ \left( \ft{25}{4} + \ft{21}{2}\cos^2\alpha + \ft{35}{4}\cos^4\alpha \right) \zeta_6 +\sin^2\alpha  \, \zeta_3^2 \right]
\nonumber\\
& + 256 g^{10} \cos^2\alpha \Big[ \left( \ft{1225}{4} + \ft{1061}{2}\cos^2\alpha + 525 \cos^4\alpha + \ft{875}{3}\cos^6\alpha \right) \zeta_8
\nonumber\\
&\qquad\qquad\qquad + 4 \cos^2\alpha \sin^2\alpha \, \zeta_2 \zeta_3^2  + 10 \sin^2 \alpha \, \zeta_3 \zeta_5
\Big]
\nonumber\\
& - 512 g^{12} \cos^2\alpha
\Big[ \left( \ft{1323}{8} + 297 \cos^2\alpha + \ft{12639}{40}\cos^4\alpha + \ft{231}{10}\cos^6\alpha + \ft{77}{8}\cos^8\alpha \right)  \zeta_{10}
\nonumber\\
&\qquad\qquad\qquad
+
\left( \ft{81}{10} + 27\cos^2\alpha \right) \cos^2\alpha\sin^2\alpha \, \zeta_4 \zeta_3^2
+
40 \cos^2\alpha \sin^2\alpha \, \zeta_2 \zeta_3 \zeta_5
\nonumber\\
&\qquad\qquad\qquad
+
\ft{51}{2} \sin^2\alpha\, \zeta_5^2 + \ft{105}{2} \sin^2\alpha \, \zeta_3 \zeta_7
\Big]
+ O (g^{14})
\, . \nonumber
\end{align}


\begin{thebibliography}{10}

\bibitem{Caron-Huot:2021usw}
S.~Caron-Huot and F.~Coronado, \emph{{Ten dimensional symmetry of $ \mathcal{N}
  $ = 4 SYM correlators}},
  \href{http://dx.doi.org/10.1007/JHEP03(2022)151}{\emph{JHEP} {\bf 03} (2022)
  151}, [\href{https://arxiv.org/abs/2106.03892}{{\tt 2106.03892}}].

\bibitem{Bork:2022vat}
L.~V. Bork, N.~B. Muzhichkov and E.~S. Sozinov, \emph{{Infrared properties of
  five-point massive amplitudes in $ \mathcal{N} $ = 4 SYM on the Coulomb
  branch}}, \href{http://dx.doi.org/10.1007/JHEP08(2022)173}{\emph{JHEP} {\bf
  08} (2022) 173}, [\href{https://arxiv.org/abs/2201.08762}{{\tt 2201.08762}}].

\bibitem{Belitsky:2022itf}
A.~V. Belitsky, L.~V. Bork, A.~F. Pikelner and V.~A. Smirnov, \emph{{Exact Off
  Shell Sudakov Form Factor in N=4 Supersymmetric Yang-Mills Theory}},
  \href{http://dx.doi.org/10.1103/PhysRevLett.130.091605}{\emph{Phys. Rev.
  Lett.} {\bf 130} (2023) 091605},
  [\href{https://arxiv.org/abs/2209.09263}{{\tt 2209.09263}}].

\bibitem{Belitsky:2023ssv}
A.~V. Belitsky, L.~V. Bork and V.~A. Smirnov, \emph{{Off-shell form factor in $
  \mathcal{N} $=4 sYM at three loops}},
  \href{http://dx.doi.org/10.1007/JHEP11(2023)111}{\emph{JHEP} {\bf 11} (2023)
  111}, [\href{https://arxiv.org/abs/2306.16859}{{\tt 2306.16859}}].

\bibitem{Belitsky:2024agy}
A.~V. Belitsky, L.~V. Bork, J.~M. Grumski-Flores and V.~A. Smirnov,
  \emph{{Three-leg form factor on Coulomb branch}},
  \href{http://dx.doi.org/10.1007/JHEP11(2024)169}{\emph{JHEP} {\bf 11} (2024)
  169}, [\href{https://arxiv.org/abs/2402.18475}{{\tt 2402.18475}}].

\bibitem{Belitsky:2024dcf}
A.~V. Belitsky and L.~V. Bork, \emph{{Off-shell minimal form factors}},
  \href{http://dx.doi.org/10.1007/JHEP07(2025)231}{\emph{JHEP} {\bf 07} (2025)
  231}, [\href{https://arxiv.org/abs/2411.16941}{{\tt 2411.16941}}].

\bibitem{Belitsky:2025bgb}
A.~V. Belitsky, L.~V. Bork, R.~N. Lee, A.~I. Onishchenko and V.~A. Smirnov,
  \emph{{Five W-boson amplitude is equal to near-null decagon}},
  \href{http://dx.doi.org/10.1103/5sf8-fhmt}{\emph{Phys. Rev. D} {\bf 113}
  (2026) 086003}, [\href{https://arxiv.org/abs/2510.16471}{{\tt 2510.16471}}].

\bibitem{Alday:2009zm}
L.~F. Alday, J.~M. Henn, J.~Plefka and T.~Schuster, \emph{{Scattering into the
  fifth dimension of N=4 super Yang-Mills}},
  \href{http://dx.doi.org/10.1007/JHEP01(2010)077}{\emph{JHEP} {\bf 01} (2010)
  077}, [\href{https://arxiv.org/abs/0908.0684}{{\tt 0908.0684}}].

\bibitem{Maldacena:1997re}
J.~M. Maldacena, \emph{{The Large $N$ limit of superconformal field theories
  and supergravity}},
  \href{http://dx.doi.org/10.4310/ATMP.1998.v2.n2.a1}{\emph{Adv. Theor. Math.
  Phys.} {\bf 2} (1998) 231--252},
  [\href{https://arxiv.org/abs/hep-th/9711200}{{\tt hep-th/9711200}}].

\bibitem{Alday:2007hr}
L.~F. Alday and J.~M. Maldacena, \emph{{Gluon scattering amplitudes at strong
  coupling}},
  \href{http://dx.doi.org/10.1088/1126-6708/2007/06/064}{\emph{JHEP} {\bf 06}
  (2007) 064}, [\href{https://arxiv.org/abs/0705.0303}{{\tt 0705.0303}}].

\bibitem{Alday:2007he}
L.~F. Alday and J.~Maldacena, \emph{{Comments on gluon scattering amplitudes
  via AdS/CFT}},
  \href{http://dx.doi.org/10.1088/1126-6708/2007/11/068}{\emph{JHEP} {\bf 11}
  (2007) 068}, [\href{https://arxiv.org/abs/0710.1060}{{\tt 0710.1060}}].

\bibitem{BBLOS2026}
A.~V. Belitsky, L.~V. Bork, R.~N. Lee, A.~I. Onishchenko and V.~A. Smirnov,
  \emph{{Five legs @ three loops: II. Integrals}},
  \href{https://arxiv.org/abs/(to appear)}{{\tt (to appear)}}.

\bibitem{Bern:1994zx}
Z.~Bern, L.~J. Dixon, D.~C. Dunbar and D.~A. Kosower, \emph{{One loop n point
  gauge theory amplitudes, unitarity and collinear limits}},
  \href{http://dx.doi.org/10.1016/0550-3213(94)90179-1}{\emph{Nucl. Phys. B}
  {\bf 425} (1994) 217--260}, [\href{https://arxiv.org/abs/hep-ph/9403226}{{\tt
  hep-ph/9403226}}].

\bibitem{Bern:1994cg}
Z.~Bern, L.~J. Dixon, D.~C. Dunbar and D.~A. Kosower, \emph{{Fusing gauge
  theory tree amplitudes into loop amplitudes}},
  \href{http://dx.doi.org/10.1016/0550-3213(94)00488-Z}{\emph{Nucl. Phys. B}
  {\bf 435} (1995) 59--101}, [\href{https://arxiv.org/abs/hep-ph/9409265}{{\tt
  hep-ph/9409265}}].

\bibitem{Bern:2004cz}
Z.~Bern, L.~J. Dixon and D.~A. Kosower, \emph{{Two-loop g ---\ensuremath{>} gg
  splitting amplitudes in QCD}},
  \href{http://dx.doi.org/10.1088/1126-6708/2004/08/012}{\emph{JHEP} {\bf 08}
  (2004) 012}, [\href{https://arxiv.org/abs/hep-ph/0404293}{{\tt
  hep-ph/0404293}}].

\bibitem{Bern:1997sc}
Z.~Bern, L.~J. Dixon and D.~A. Kosower, \emph{{One loop amplitudes for e+ e- to
  four partons}},
  \href{http://dx.doi.org/10.1016/S0550-3213(97)00703-7}{\emph{Nucl. Phys. B}
  {\bf 513} (1998) 3--86}, [\href{https://arxiv.org/abs/hep-ph/9708239}{{\tt
  hep-ph/9708239}}].

\bibitem{Britto:2004nc}
R.~Britto, F.~Cachazo and B.~Feng, \emph{{Generalized unitarity and one-loop
  amplitudes in N=4 super-Yang-Mills}},
  \href{http://dx.doi.org/10.1016/j.nuclphysb.2005.07.014}{\emph{Nucl. Phys. B}
  {\bf 725} (2005) 275--305}, [\href{https://arxiv.org/abs/hep-th/0412103}{{\tt
  hep-th/0412103}}].

\bibitem{Cachazo:2008vp}
F.~Cachazo, \emph{{Sharpening The Leading Singularity}},
  \href{https://arxiv.org/abs/0803.1988}{{\tt 0803.1988}}.

\bibitem{Selivanov:1999ie}
K.~G. Selivanov, \emph{{An Infinite set of tree amplitudes in
  Higgs-Yang-Mills}},
  \href{http://dx.doi.org/10.1016/S0370-2693(99)00760-1}{\emph{Phys. Lett. B}
  {\bf 460} (1999) 116--118}, [\href{https://arxiv.org/abs/hep-th/9906001}{{\tt
  hep-th/9906001}}].

\bibitem{Bern:2010qa}
Z.~Bern, J.~J. Carrasco, T.~Dennen, Y.-t. Huang and H.~Ita, \emph{{Generalized
  Unitarity and Six-Dimensional Helicity}},
  \href{http://dx.doi.org/10.1103/PhysRevD.83.085022}{\emph{Phys. Rev. D} {\bf
  83} (2011) 085022}, [\href{https://arxiv.org/abs/1010.0494}{{\tt
  1010.0494}}].

\bibitem{Boels:2010mj}
R.~H. Boels, \emph{{No triangles on the moduli space of maximally
  supersymmetric gauge theory}},
  \href{http://dx.doi.org/10.1007/JHEP05(2010)046}{\emph{JHEP} {\bf 05} (2010)
  046}, [\href{https://arxiv.org/abs/1003.2989}{{\tt 1003.2989}}].

\bibitem{Craig:2011ws}
N.~Craig, H.~Elvang, M.~Kiermaier and T.~Slatyer, \emph{{Massive amplitudes on
  the Coulomb branch of N=4 SYM}},
  \href{http://dx.doi.org/10.1007/JHEP12(2011)097}{\emph{JHEP} {\bf 12} (2011)
  097}, [\href{https://arxiv.org/abs/1104.2050}{{\tt 1104.2050}}].

\bibitem{Caron-Huot:2010nes}
S.~Caron-Huot and D.~O'Connell, \emph{{Spinor Helicity and Dual Conformal
  Symmetry in Ten Dimensions}},
  \href{http://dx.doi.org/10.1007/JHEP08(2011)014}{\emph{JHEP} {\bf 08} (2011)
  014}, [\href{https://arxiv.org/abs/1010.5487}{{\tt 1010.5487}}].

\bibitem{Cheung:2009dc}
C.~Cheung and D.~O'Connell, \emph{{Amplitudes and Spinor-Helicity in Six
  Dimensions}},
  \href{http://dx.doi.org/10.1088/1126-6708/2009/07/075}{\emph{JHEP} {\bf 07}
  (2009) 075}, [\href{https://arxiv.org/abs/0902.0981}{{\tt 0902.0981}}].

\bibitem{Dennen:2009vk}
T.~Dennen, Y.-t. Huang and W.~Siegel, \emph{{Supertwistor space for 6D maximal
  super Yang-Mills}},
  \href{http://dx.doi.org/10.1007/JHEP04(2010)127}{\emph{JHEP} {\bf 04} (2010)
  127}, [\href{https://arxiv.org/abs/0910.2688}{{\tt 0910.2688}}].

\bibitem{Plefka:2014fta}
J.~Plefka, T.~Schuster and V.~Verschinin, \emph{{From Six to Four and More:
  Massless and Massive Maximal Super Yang-Mills Amplitudes in 6d and 4d and
  their Hidden Symmetries}},
  \href{http://dx.doi.org/10.1007/JHEP01(2015)098}{\emph{JHEP} {\bf 01} (2015)
  098}, [\href{https://arxiv.org/abs/1405.7248}{{\tt 1405.7248}}].

\bibitem{Huang:2011um}
Y.-t. Huang, \emph{{Non-Chiral S-Matrix of N=4 Super Yang-Mills}},
  \href{https://arxiv.org/abs/1104.2021}{{\tt 1104.2021}}.

\bibitem{Belitsky:2024rwv}
A.~V. Belitsky, \emph{{Collinear anatomy}},
  \href{http://dx.doi.org/10.1007/JHEP05(2025)117}{\emph{JHEP} {\bf 05} (2025)
  117}, [\href{https://arxiv.org/abs/2412.11886}{{\tt 2412.11886}}].

\bibitem{Belitsky:2025vfc}
A.~V. Belitsky, \emph{{Towards six W-boson amplitude at two loops}},
  \href{https://arxiv.org/abs/2511.20828}{{\tt 2511.20828}}.

\bibitem{Bern:2002tk}
Z.~Bern, A.~De~Freitas and L.~J. Dixon, \emph{{Two loop helicity amplitudes for
  gluon-gluon scattering in QCD and supersymmetric Yang-Mills theory}},
  \href{http://dx.doi.org/10.1088/1126-6708/2002/03/018}{\emph{JHEP} {\bf 03}
  (2002) 018}, [\href{https://arxiv.org/abs/hep-ph/0201161}{{\tt
  hep-ph/0201161}}].

\bibitem{Spradlin:2008uu}
M.~Spradlin, A.~Volovich and C.~Wen, \emph{{Three-Loop Leading Singularities
  and BDS Ansatz for Five Particles}},
  \href{http://dx.doi.org/10.1103/PhysRevD.78.085025}{\emph{Phys. Rev. D} {\bf
  78} (2008) 085025}, [\href{https://arxiv.org/abs/0808.1054}{{\tt
  0808.1054}}].

\bibitem{Ambrosio:2013pba}
R.~G. Ambrosio, B.~Eden, T.~Goddard, P.~Heslop and C.~Taylor, \emph{{Local
  integrands for the five-point amplitude in planar N=4 SYM up to five loops}},
  \href{http://dx.doi.org/10.1007/JHEP01(2015)116}{\emph{JHEP} {\bf 01} (2015)
  116}, [\href{https://arxiv.org/abs/1312.1163}{{\tt 1312.1163}}].

\bibitem{Coronado:2018cxj}
F.~Coronado, \emph{{Bootstrapping the Simplest Correlator in Planar $\mathcal N
  = 4$ Supersymmetric Yang-Mills Theory to All Loops}},
  \href{http://dx.doi.org/10.1103/PhysRevLett.124.171601}{\emph{Phys. Rev.
  Lett.} {\bf 124} (2020) 171601},
  [\href{https://arxiv.org/abs/1811.03282}{{\tt 1811.03282}}].

\bibitem{Belitsky:2019fan}
A.~V. Belitsky and G.~P. Korchemsky, \emph{{Exact null octagon}},
  \href{http://dx.doi.org/10.1007/JHEP05(2020)070}{\emph{JHEP} {\bf 05} (2020)
  070}, [\href{https://arxiv.org/abs/1907.13131}{{\tt 1907.13131}}].

\bibitem{Bercini:2020msp}
C.~Bercini, V.~Gon{\c{c}}alves and P.~Vieira, \emph{{Light-Cone Bootstrap of
  Higher Point Functions and Wilson Loop Duality}},
  \href{http://dx.doi.org/10.1103/PhysRevLett.126.121603}{\emph{Phys. Rev.
  Lett.} {\bf 126} (2021) 121603},
  [\href{https://arxiv.org/abs/2008.10407}{{\tt 2008.10407}}].

\bibitem{Basso:2020xts}
B.~Basso, L.~J. Dixon and G.~Papathanasiou, \emph{{Origin of the Six-Gluon
  Amplitude in Planar $N=4$ Supersymmetric Yang-Mills Theory}},
  \href{http://dx.doi.org/10.1103/PhysRevLett.124.161603}{\emph{Phys. Rev.
  Lett.} {\bf 124} (2020) 161603},
  [\href{https://arxiv.org/abs/2001.05460}{{\tt 2001.05460}}].

\bibitem{Bercini:2024pya}
C.~Bercini, B.~Fernandes and V.~Gon{\c{c}}alves, \emph{{Two-loop five-point
  integrals: light, heavy and large-spin correlators}},
  \href{http://dx.doi.org/10.1007/JHEP10(2024)242}{\emph{JHEP} {\bf 10} (2024)
  242}, [\href{https://arxiv.org/abs/2401.06099}{{\tt 2401.06099}}].

\bibitem{Basso:2015zoa}
B.~Basso, S.~Komatsu and P.~Vieira, \emph{{Structure Constants and Integrable
  Bootstrap in Planar N=4 SYM Theory}},
  \href{https://arxiv.org/abs/1505.06745}{{\tt 1505.06745}}.

\bibitem{Fleury:2016ykk}
T.~Fleury and S.~Komatsu, \emph{{Hexagonalization of Correlation Functions}},
  \href{http://dx.doi.org/10.1007/JHEP01(2017)130}{\emph{JHEP} {\bf 01} (2017)
  130}, [\href{https://arxiv.org/abs/1611.05577}{{\tt 1611.05577}}].

\bibitem{Fleury:2017eph}
T.~Fleury and S.~Komatsu, \emph{{Hexagonalization of Correlation Functions II:
  Two-Particle Contributions}},
  \href{http://dx.doi.org/10.1007/JHEP02(2018)177}{\emph{JHEP} {\bf 02} (2018)
  177}, [\href{https://arxiv.org/abs/1711.05327}{{\tt 1711.05327}}].

\bibitem{Sterman:2002qn}
G.~F. Sterman and M.~E. Tejeda-Yeomans, \emph{{Multiloop amplitudes and
  resummation}},
  \href{http://dx.doi.org/10.1016/S0370-2693(02)03100-3}{\emph{Phys. Lett. B}
  {\bf 552} (2003) 48--56}, [\href{https://arxiv.org/abs/hep-ph/0210130}{{\tt
  hep-ph/0210130}}].

\bibitem{Aybat:2006mz}
S.~M. Aybat, L.~J. Dixon and G.~F. Sterman, \emph{{The Two-loop soft anomalous
  dimension matrix and resummation at next-to-next-to leading pole}},
  \href{http://dx.doi.org/10.1103/PhysRevD.74.074004}{\emph{Phys. Rev. D} {\bf
  74} (2006) 074004}, [\href{https://arxiv.org/abs/hep-ph/0607309}{{\tt
  hep-ph/0607309}}].

\bibitem{Dixon:2008gr}
L.~J. Dixon, L.~Magnea and G.~F. Sterman, \emph{{Universal structure of
  subleading infrared poles in gauge theory amplitudes}},
  \href{http://dx.doi.org/10.1088/1126-6708/2008/08/022}{\emph{JHEP} {\bf 08}
  (2008) 022}, [\href{https://arxiv.org/abs/0805.3515}{{\tt 0805.3515}}].

\bibitem{Becher:2009qa}
T.~Becher and M.~Neubert, \emph{{On the Structure of Infrared Singularities of
  Gauge-Theory Amplitudes}},
  \href{http://dx.doi.org/10.1088/1126-6708/2009/06/081}{\emph{JHEP} {\bf 06}
  (2009) 081}, [\href{https://arxiv.org/abs/0903.1126}{{\tt 0903.1126}}].

\bibitem{Anastasiou:2003kj}
C.~Anastasiou, Z.~Bern, L.~J. Dixon and D.~A. Kosower, \emph{{Planar amplitudes
  in maximally supersymmetric Yang-Mills theory}},
  \href{http://dx.doi.org/10.1103/PhysRevLett.91.251602}{\emph{Phys. Rev.
  Lett.} {\bf 91} (2003) 251602},
  [\href{https://arxiv.org/abs/hep-th/0309040}{{\tt hep-th/0309040}}].

\bibitem{Bern:2005iz}
Z.~Bern, L.~J. Dixon and V.~A. Smirnov, \emph{{Iteration of planar amplitudes
  in maximally supersymmetric Yang-Mills theory at three loops and beyond}},
  \href{http://dx.doi.org/10.1103/PhysRevD.72.085001}{\emph{Phys. Rev. D} {\bf
  72} (2005) 085001}, [\href{https://arxiv.org/abs/hep-th/0505205}{{\tt
  hep-th/0505205}}].

\bibitem{Belitsky:2025sin}
A.~V. Belitsky and V.~A. Smirnov, \emph{{Tropical regions of near mass-shell
  pentabox}}, \href{http://dx.doi.org/10.1103/hn9g-ccd8}{\emph{Phys. Rev. D}
  {\bf 113} (2026) 045019}, [\href{https://arxiv.org/abs/2508.14298}{{\tt
  2508.14298}}].

\bibitem{Beneke:1997zp}
M.~Beneke and V.~A. Smirnov, \emph{{Asymptotic expansion of Feynman integrals
  near threshold}},
  \href{http://dx.doi.org/10.1016/S0550-3213(98)00138-2}{\emph{Nucl. Phys. B}
  {\bf 522} (1998) 321--344}, [\href{https://arxiv.org/abs/hep-ph/9711391}{{\tt
  hep-ph/9711391}}].

\bibitem{Bork:2025ztu}
L.~V. Bork, R.~N. Lee and A.~I. Onishchenko, \emph{{Method of regions for dual
  conformal integrals}},
  \href{http://dx.doi.org/10.1007/JHEP12(2025)107}{\emph{JHEP} {\bf 12} (2025)
  107}, [\href{https://arxiv.org/abs/2509.12056}{{\tt 2509.12056}}].

\bibitem{Panzer:2014gra}
E.~Panzer, \emph{{On hyperlogarithms and Feynman integrals with divergences and
  many scales}}, \href{http://dx.doi.org/10.1007/JHEP03(2014)071}{\emph{JHEP}
  {\bf 03} (2014) 071}, [\href{https://arxiv.org/abs/1401.4361}{{\tt
  1401.4361}}].

\end{thebibliography}
\end{document}